\documentclass[12pt]{article}
\usepackage{graphicx}
\usepackage{subcaption} 
\usepackage{amsmath,amsfonts,amssymb,stmaryrd,latexsym}
\usepackage{url,comment}    
\usepackage{hyperref}
\usepackage{xcolor}
\hypersetup{
    colorlinks=true,       
    linkcolor=purple,   
    citecolor=blue,        
    filecolor=magenta,      
    urlcolor=black           
}
\usepackage{cite}
\usepackage{dcolumn}
\usepackage{bm}
\usepackage{slashed}
\usepackage[compat=1.0.0]{tikz-feynman}

\newcommand{\be}{\begin{equation}}
\newcommand{\ee}{\end{equation}}
\newcommand{\bea}{\begin{eqnarray}}
\newcommand{\eea}{\end{eqnarray}}
\newcommand{\bear}{\begin{eqnarray}}
\newcommand{\eear}{\end{eqnarray}}
\newcommand{\ba}{\begin{array}}
\newcommand{\ea}{\end{array}}

\newcommand\identity{1\kern-0.25em\text{l}}

\begin{document}

\baselineskip=18pt \pagestyle{plain} \setcounter{page}{1}

\vspace*{-1cm}

\noindent \makebox[11.9cm][l]{\small \hspace*{-.2cm} }{\small Fermilab-PUB-24-0958-T}  \\  [-1mm]
\noindent \makebox[11.9cm][l]{\small \hspace*{-.2cm} }{\small MITP/23-035}  \\  [-1mm]

\begin{center}

{\large \bf    
Quark-universal $U(1)$ breaking scalar at the LHC
} \\ [11mm]

{\normalsize \bf Lorin Armbruster$^\star$, Bogdan A. Dobrescu$^\diamond$, Felix Yu$^\star$ \\ [6mm]
{\small {\it  $^\star$ PRISMA+ Cluster of Excellence \& Mainz Institute for
  Theoretical Physics, \\ Johannes Gutenberg University, 55099 Mainz,
  Germany}}\\ [3mm]
{\small {\it $^\diamond$ Particle Theory Department, Fermilab, Batavia, IL 60510, USA     }}\\  [3mm]
}

\center{June 6, 2025}
\end{center}

\vspace*{0.2cm}

\begin{abstract}
If the quarks or leptons are charged under a new $U(1)$ gauge
symmetry, then besides a $Z'$ boson there must exist at least one new
boson whose decay products include Standard Model particles. In the
case of a minimal symmetry breaking sector, that new boson is a scalar
$\phi$ that couples to the $Z'$ boson as well as to the new
fermions required to cancel the $U(1)$ gauge anomalies.  The scalar
may be produced at the Large Hadron Collider (LHC) in association with a $Z'$ boson, or
through $Z'$ boson fusion, while its decays are typically into four
jets or two photons. We analyze in detail the case where the $Z'$
boson is leptophobic, and all the quarks have the same charge under
the new $U(1)$.  If $\phi$ mixes with the Standard Model Higgs
boson, then the new scalar can also be produced via gluon fusion, and the
discovery mode is likely to be a diphoton resonance.
\end{abstract}

\vspace*{0.1cm}

\newpage

\renewcommand{\contentsname}{\normalsize\large Contents}
{\small
\hypersetup{linktocpage} 
\tableofcontents
\hypersetup{linkcolor=red} 
}

\vspace*{1.1cm}

\section{Introduction} \setcounter{equation}{0}
\label{sec:intro}

In the Standard Model (SM), the Higgs field plays a central role as
the source of electroweak symmetry breaking via its vacuum expectation
value (vev), generating the masses of the electroweak gauge bosons and
the charged SM fermions.  In $U(1)$ gauge extensions 
of the SM, such as gauging a linear combination of baryon and lepton global
number~\cite{Pais:1973mi, Carone:1994aa, Bailey:1994qv, Carone:1995pu,
  Aranda:1998fr, Carena:2004xs, FileviezPerez:2011pt, Duerr:2013dza,
  Dobrescu:2013cmh, Dobrescu:2014fca, Dobrescu:2015asa,
  Dobrescu:2017sue, Michaels:2020fzj, Dobrescu:2021vak,
  Kivel:2022rzr}, generic charge choices require new
fermions to cancel the gauge anomalies (``anomalons'').  In addition, introducing a scalar field $\Phi$ that spontaneously breaks the $U(1)$ gauge symmetry affords a renormalizable description to generate the $Z'$ mass as well as
anomalon masses, avoiding phenomenological constraints from measurements of the 125~GeV Higgs boson~\cite{Michaels:2020fzj,  Dobrescu:2021vak}.  

In this paper we study the properties and Large Hadron Collider (LHC) phenomenology of the new scalar particle ($\phi$, the radial degree of freedom of $\Phi$) responsible for the $Z'$ mass.
Generically, $\phi$  production at the LHC predominantly occurs in association with a $Z'$ boson, and through $Z'$-mediated vector boson fusion.
We focus on the predictions of the minimal renormalizable $U(1)_B$ model, in which the $Z'$ boson has flavor-universal couplings to the SM quarks, 
and the anomalons are color singlets but carry chiral electroweak and $U(1)_B$ charges \cite{Duerr:2013dza,  Dobrescu:2014fca, Dobrescu:2015asa}. 
Since the anomalons are not colored, the new scalar would not be produced through gluon fusion at the LHC except via a model-dependent mixing with the 125~GeV Higgs boson.  
This contrasts other work on the Higgs sector of gauged baryon number symmetry~\cite{Duerr:2017whl}, which relied on a colored anomalon content. 

On the decay side, the scalar associated with new $U(1)$ chiral symmetries
decays to $Z'$ pairs as a result of spontaneous symmetry breaking, as
well as to pairs of $U(1)$ chiral fermions proportional to Yukawa
couplings.  If decays to fermions are kinematically disallowed, the
anomalons mediate loop-induced decays of $\phi$  to any and
all gauge bosons under which the anomalons are charged.  These loop
interactions are nondecoupling as a result of Higgs low-energy
theorems, and provide a characteristic pattern of Higgs decays in
chiral theories compared to theories where fermions also have 
vectorlike masses~\cite{Shifman:1979eb, Kumar:2012ww}.  In our study
of a quark-universal $U(1)_B$ gauge symmetry with anomalons carrying
SM electroweak charges, the $\phi$ scalar acquires 1-loop couplings to pairs of gauge bosons, most importantly $\gamma \gamma$, $Z' \gamma$, $Z \gamma$, $Z'Z$, $ZZ$, and $WW$.

Mass mixing between $\phi$ and the SM Higgs boson leads to an important interplay
between the phenomenology of the Higgs boson and of the new physical scalar, denoted by $\varphi$.  For the
quark-universal $U(1)_B$ model, since $\phi$ has
intrinsic decays to $\gamma \gamma$, $Z \gamma$, $ZZ$, and $WW$, its effects
on the 125~GeV SM-like Higgs boson are not solely captured by the
typical interpretation of a mixing angle with a SM gauge singlet
scalar.  This pattern of coupling modifications on the observed
125~GeV Higgs boson offers a new benchmark to expand the interpretation
power of the current suite of coupling measurements by the ATLAS and
CMS experiments~\cite{CMS:2022dwd, ATLAS:2022vkf, ATLAS:2024fkg}.
Crucially, the imprints of the SM Higgs phenomenology on the new scalar dictate interference effects sensitive to the mixing
angle parameter and the individual chiral symmetries seen by the
unmixed Higgses.  A related work that studies the anomalon content
from the point of view of dark matter is~\cite{Butterworth:2024eyr}.

The outline of this paper is as follows.  In Section~\ref{sec:theory}, we
analyze the quark-universal $U(1)_B$ gauge symmetry, the anomalon
fields, and the spontaneous breaking of $U(1)_B$ symmetry by the complex scalar field $\Phi$.  In particular, we discuss the production
and decay modes for the radial scalar mode of $\Phi$.
In Section~\ref{sec:phiHiggs}, we reanalyze the phenomenology
when $\phi$ mixes with the SM Higgs boson, taking into account the
current constraints from measurements of the 125~GeV Higgs boson.  We
conclude in Section~\ref{sec:conclusions}.

\section{Quark-universal $U(1)_B$ gauge symmetry}
\label{sec:theory}

A quark-universal $U(1)_B$ gauge symmetry is a chiral gauge extension
of the Standard Model gauge sector, since the global baryon number
symmetry of the SM is anomalous with respect to $SU(2)_L$ and
$U(1)_Y$.  Hence, renormalizability of the theory requires adding new
fermions (``anomalons'') to cancel the mixed $SU(2)_L^2 \times U(1)_B$
and $U(1)_Y^2 \times U(1)_B$ anomalies, and also must ensure no new
gauge anomalies are introduced~\cite{Preskill:1990fr}.  Furthermore,
we introduce a scalar field $\Phi$ that acquires a vacuum expectation
value (vev) to spontaneously break the $U(1)_B$ symmetry and generate
masses for the $Z'$ boson and anomalons, which is also a minimal
construction consistent with renormalizability and perturbative
unitarity~\cite{Appelquist:1987cf, Kribs:2022gri}.  Hence, a
renormalizable theory for a $U(1)_B$ gauge symmetry with
quark-universal couplings is the fermion set~\cite{Duerr:2013dza,
  Dobrescu:2014fca, Dobrescu:2015asa}
\bear
L_L \sim (2, -1/2, -1), \quad E_L \sim (1, -1, 2), \quad N_L \sim (1, 0, 2),
\nonumber \\
L_R \sim (2, -1/2, 2), \quad E_R \sim (1, -1, -1), \quad N_R \sim (1, 0, -1) \ ,
\label{eq:anomalons}
\eear
where representations under $SU(2)_L \times U(1)_Y \times U(1)_B$ are
given, and we also introduce a complex scalar $\Phi$ that is a SM
gauge singlet and carries $U(1)_B$ charge $z_\Phi=+3$.  The $N_L$ and $N_R$
fields are necessary to cancel the $U(1)_B$-gravitational anomaly and
$U(1)_B^3$ anomalies~\cite{Dobrescu:2013cmh, Dobrescu:2014fca,
  Michaels:2020fzj, Dobrescu:2021vak}.  An additional feature of the
above matter content is that the kinetic mixing between the SM $Z$
boson and the $Z'$ boson is only generated at 1-loop and is
logarithmic below the mass scale of the
anomalons~\cite{Dobrescu:2021vak}.  A tree-level kinetic mixing can be
forbidden given the theory is embedded in a non-Abelian ultraviolet
completion and remains vanishing until the scale of the anomalon
masses since the fermion content of the model satisfies the trace
condition~\cite{Dobrescu:2021vak}.  An important distinction from
other studies of gauged baryon number~\cite{FileviezPerez:2010gw,
  FileviezPerez:2011pt, Duerr:2017whl} is that our anomalon sector is
uncharged under $SU(3)_c$ symmetry, which dramatically alters the
collider phenomenology for anomalon production as well as the radial
mode of $\Phi$.

Having established the matter content for the chiral $U(1)_B$ gauge
symmetry, we now discuss the Lagrangian interactions, beginning with
the scalar field $\Phi$.

\subsection{Scalar responsible for $U(1)_B$ breaking and anomalon masses}
\label{subsec:phi}

A renormalizable realization of the $U(1)_B$ spontaneous
symmetry breaking is the Lagrangian,
\be
\mathcal{L} = - \frac{1}{4} Z'_{\mu \nu} Z'^{\mu \nu} + |D_\mu \Phi|^2
 - \mu_\Phi^2 |\Phi|^2 - \lambda_\Phi |\Phi|^4 \ ,
\ee
where the covariant derivative is 
\be
D_\mu = \partial_\mu - 3 i \frac{g_{_B}}{2} Z'_\mu  ~~,
\ee
and $\mu_\Phi^2 < 0$ triggers spontaneous breaking of $U(1)_B$ via the vev $v' = \sqrt{
  |\mu_\Phi^2| / \lambda_\Phi}$, leaving the radial mode $\Phi = (v' +
\phi) / \sqrt{2}$ (in unitary gauge).  The $U(1)_B$ interactions
for SM quarks are
\be
\mathcal{L}_q = \frac{g_{_B}}{6} Z^\prime_{\mu} \sum_q \, 
\overline q \gamma^\mu q  ~~,
\label{eq:quark-couplings}
\ee
where $q$ represents the six SM quark flavors, and the normalization accounts for  the baryon charge ($+1/3$) of the SM quarks.

Discovering the physical real scalar $\phi$ at colliders is a
necessary requirement to establish the spontaneous symmetry breaking
nature of the $Z'$ boson.  Moreover, the $\phi$ scalar cannot be
too heavy compared to the $Z'$ boson, since the $\phi$ boson
unitarizes longitudinal $Z'$ scattering~\cite{Kribs:2022gri,
  Lee:1977eg}.  The $Z'$ mass is $M_{Z'} = 3 \frac{g_{_B}}{2} v'$.

The anomalons have the Yukawa Lagrangian
\bear
\mathcal{L}_\mathrm{Yuk} = \hspace*{-0.5cm}
    && - \, y_L \bar{L}_L \Phi^* L_R - y_E \bar{E}_L \Phi E_R - y_N \bar{N}_L \Phi N_R  \nonumber \\
    &&  - \, y_1 \bar{L}_L H E_R - y_2 \bar{L}_R H E_L - y_3 \bar{L}_L \tilde{H} N_R - y_4 \bar{L}_R \tilde{H} N_L + \mathrm{h.c.} \ ,
    \label{eq:lagrangian:yuk}
\eear
where $H$ is the complex scalar Higgs doublet of the SM.  The first
line of Eq.~(\ref{eq:lagrangian:yuk}) reflects the chiral nature of
the anomalons under the $U(1)_B$ symmetry and the second line reflects
the chiral nature under the SM gauge symmetries.  We can use global
$U(1)$ phase transformations on the fields to ensure that all the
Yukawa couplings in (\ref{eq:lagrangian:yuk}) are real.

Note that Yukawa couplings of anomalons to SM leptons are forbidden by
the $U(1)_B$ gauge symmetry. Even after $U(1)_B$ breaking, a remnant
$\mathbb{Z}_3$  symmetry (a consequence of the $U(1)_B$ charge of $\Phi$, $z_\Phi=3$) 
still forbids such couplings. In extended models (not
analyzed here) that include scalars with vevs that carry $U(1)_B$
charge 1 or 2, Yukawa couplings of anomalons to SM leptons would be
allowed (see \cite{Bernreuther:2023uxh} for a study of their possible
effects).

Since both $\Phi$ and $H$ acquire vevs ($v'$ and $v_h \approx 246$
GeV, respectively), the anomalon mass eigenstates are admixtures of
the Weyl fermions in Eq.~(\ref{eq:anomalons}), with two Dirac fermions
$E_1$ and $E_2$ of electric charge $-1$ and two Dirac fermions $N_1$
and $N_2$ that are electrically neutral.  The $E_1$ and $E_2$ masses
are
\bear
\hspace*{-0.3cm}
M_{1,\, 2}^2 = \dfrac{1}{4} \left( v_h^2 (y_1^2 + y_2^2)  + v'^2 (y_L^2 + y_E^2)  \rule{0mm}{4mm}  \right) 
\!
\left( 1 \mp \sqrt{ 1 - \dfrac{4 (v_h^2 \, y_1 y_2 - v'^2 \, y_L y_E)^2}{ \left( v_h^2 (y_1^2 + y_2^2) + v'^2 (y_L^2 + y_E^2) \rule{0mm}{3.6mm} \right)^2 }} \, \right) .
\eear
The interaction and mass eigenstates for the charged anomalons are related by separate rotations for the left- and right-handed fields:
\bear
\begin{split}
& \left( \ba{c}   L^{\! ^-}_{L}  \\  E_{L}   \ea  \right)   = 
\left( \ba{cc}  \cos\theta   &  \sin\theta   \\ [2mm]  -\sin\theta   &    \cos\theta    \ea  \right)  
\left( \ba{c}   E_{2_L}  \\  E_{1_L}    \ea  \right)    ~~\, ,
\\[2mm]
& \left( \ba{c}    L^{\! ^-}_{R}  \\ E_{R}   \ea  \right)   = 
\left( \ba{cc}  \cos\chi   &  \sin\chi    \\ [2mm]  -\sin\chi   &    \cos\chi     \ea  \right)   
\left( \ba{c}   E_{2_R}  \\  E_{1_R}    \ea  \right)    ~~\, ,
 \end{split}
\eear
where the $\theta$ and $\chi$ angles satisfy
\bear
\begin{split}
\tan 2 \theta = & \dfrac{ 2 \, v_h \, v' \, ( y_L \, y_2 \, + y_E \, y_1 ) } { v'^2 \, ( y_L^2 - y_E^2) + v_h^2 \, ( y_1^2 - y_2^2) } ~~ , \\[2mm]
\tan 2 \chi = & 
    \dfrac{ 2 \, v_h \, v' \, ( y_L \, y_1 \, + y_E \, y_2 ) } { v'^2 \, ( y_L^2 - y_E^2) - v_h^2 \, ( y_1^2 - y_2^2) }
   \ .
   \end{split}
\eear

Interestingly, the Yukawa couplings in (\ref{eq:lagrangian:yuk}) can be analytically expressed in terms of the masses and mixing angles. 
To derive those relations for the charged anomalons, replace the scalars by their vevs in (\ref{eq:lagrangian:yuk}), then transform the fermions from the gauge eigenstates to mass basis. The $2\times 2$ mass matrix obtained this way, with entries proportional to linear combinations of $y_1, y_2, y_L, y_E$, is then identified with the diagonal mass matrix for $E_1$ and $E_2$. The ensuing four equations can be solved, with the result 
\bear
    \begin{split}
y_L \pm y_E = -\sqrt{2} \frac{(M_2 \pm M_1)}{v'} \cos (\theta \mp \chi) ~~ ,
\\[3mm]
y_2 \pm  y_1 = -\sqrt{2} \frac{(M_2 \mp M_1)}{v_h} \sin (\theta \pm \chi) ~~.
\label{eq:ys}
\end{split}
\eear

Having both vectorlike and chiral mass generation for the
anomalons implies the presence of off-diagonal Yukawa
couplings as well as off-diagonal  $Z$ and $Z'$ couplings in the anomalon
mass basis.  We note that the photon couplings remain diagonal, 
as a consequence of the unbroken $U(1)_{\text{em}}$
symmetry. In the mass basis, the charged anomalons have Yukawa interactions with
the $\phi$ scalar and the SM Higgs boson $h_{\text{SM}}^0$, given by
\begin{align}
\mathcal{L}_{\text{Yuk}} = 
& \, -\phi \,   \overline{E}_1 \left( 
y_{11}^\phi \, E_1 + 
y_{12}^\phi \,   E_2  \right) 
 -\phi \,   \overline{E}_2 \left( 
y_{21}^\phi \, E_1 +
y_{22}^\phi \,  E_2 
\right) \nonumber  \\ 
& - h_{\text{SM}}^0 \,  \overline{E}_1 \left( 
y_{11}^{h_{\text{SM}}} \, E_1 +
y_{12}^{h_{\text{SM}}} \,  E_2   \right) 
-h_{\text{SM}}^0 \,  \overline{E}_2  \left( 
y_{21}^{h_{\text{SM}}} \,  E_1 +
y_{22}^{h_{\text{SM}}} \,  E_2
\right) \ .
\label{eq:Ycouplings}
\end{align}
Using Eqs.~(\ref{eq:ys}), we can express the above  Yukawa couplings (times $\gamma^5$ in the case of some off-diagonal terms)  solely in terms of the mixing angles, masses, and vevs; the ones to $\phi$ are
\bear
    \begin{split}
\left\{ y_{11}^\phi \, , y_{22}^\phi  \right\} &=
\frac{1}{2 v'} \left( \mp (M_1 - M_2) \cos^2 (\theta + \chi) + (M_1 + M_2) \cos^2 (\theta - \chi)  \rule{0mm}{4mm}  \right) \ , 
\\[2mm]
\left\{  y_{12}^\phi \, , y_{21}^\phi   \right\}  &=
\frac{1}{4 v'} \left( -(M_1 - M_2) \sin 2(\theta + \chi) \pm (M_1 + M_2) \sin 2(\theta - \chi) \, \gamma^5 \rule{0mm}{4mm}  \right) \ , 
\label{eq:couplings}
\end{split}
\eear
while the ones to $h_{\text{SM}}^0$ are
\bear
    \begin{split}
\left\{  y_{11}^{h_{\text{SM}}} \, , y_{22}^{h_{\text{SM}}}  \right\}   &=
\frac{1}{2 v_h } \left( \mp (M_1 - M_2) \sin^2 (\theta + \chi) + (M_1 + M_2) \sin^2 (\theta - \chi)  \rule{0mm}{4mm}  \right) \ , 
\\[2mm]
\left\{  y_{12}^{h_{\text{SM}}} \, , y_{21}^{h_{\text{SM}}}  \right\}   &=  \frac{-v'}{v_h } \left\{  y_{12}^\phi \, , y_{21}^\phi   \right\}  \ .
\label{eq:couplings-h}
\end{split}
\eear
For later convenience, we write each coupling in (\ref{eq:couplings}) and (\ref{eq:couplings-h}) as 
\be
y_{ab}^X = y_{ab}^{X (S)} + y_{ab}^{X (P)} \gamma^5  ~~,
\label{eq:ySyP}
\ee
where $X$ here stands for $\phi$ or $h_{\text{SM}}$, and the
superscripts $(S)$ or $(P)$ denote the scalar or pseudoscalar coupling
in the given Yukawa coupling.  These Yukawa couplings play an
important role in mediating the loop-induced decays $\phi \to \gamma
\gamma$, $Z' \gamma$, and $Z \gamma$ as well as $\phi \to Z' Z$, $ZZ$,
and $W^+ W^-$. Furthermore, they lead to corrections to the SM Higgs
decay rates $h_{\text{SM}}^0 \to \gamma \gamma$ and $Z \gamma$, and
induce the nonstandard modes $h_{\text{SM}}^0 \to Z' \gamma$ and
$h_{\text{SM}}^0 \to Z' Z'$.

We remark that the off-diagonal couplings $y_{12}^\phi$ and
$y_{21}^\phi$ vanish in the limit $y_1$, $y_2 \to 0$ for finite $y_L$
and $y_E$, while an equivalent set of couplings
$y_{11}^{h_{\text{SM}}}$ and $y_{22}^{h_{\text{SM}}}$ vanish in the
limit $y_L$, $y_E \to 0$ for finite $y_1$, $y_2$.  Since $y_1$ and
$y_2$ are constrained directly by the measured 125~GeV Higgs diphoton
decay rate~\cite{CMS:2022dwd, ATLAS:2022vkf, ATLAS:2024fkg}, we will
focus on the regime $y_1$, $y_2 \ll y_L$, $y_E$ for $v' \sim v_h$,
such that the dominant terms for the anomalon masses arise from the
$U(1)_B$ breaking vev.  Note that the Yukawa couplings in
Eqs.~(\ref{eq:couplings}) and (\ref{eq:couplings-h}) correspond to
scalar states that are not mixed, and are readily amended to
incorporate a scalar mixing angle between $\phi$ and
$h_{\text{SM}}^0$, as discussed in Sec.~\ref{sec:phiHiggs}.

The Lagrangian for $Z'$ and $Z$ interactions with the charged anomalons in the mass eigenstate basis is
\begin{align}
\mathcal{L}_{Z', \, Z} = 
& \, Z'_\mu \,  \bar{E}_1 \,  \gamma^\mu  \left(  g_{11}^{Z'} \, E_1 + g_{12}^{Z'} \, E_2 \right) 
+ Z'_\mu \,  \bar{E}_2 \,  \gamma^\mu  \left(  g_{21}^{Z'}  \, E_1 +  g_{22}^{Z'}  \, E_2  \right)   
\nonumber  \\
& + Z_\mu \, \bar{E}_1 \, \gamma^\mu \left( g_{11}^{Z}  \, E_1 + g_{12}^{Z}  \, E_2 \right) 
+ Z_\mu \, \bar{E}_2 \, \gamma^\mu \left( g_{21}^{Z}  \, E_1 + g_{22}^{Z} \, E_2 \right) \ ,
\end{align}
where the vector and axial-vector couplings are encoded by the following coefficients  (times $\gamma^5$ in the case of some off-diagonal terms):
\bear
\begin{split}
g_{11}^{Z'} &= \frac{g_{_B}}{4}  \left[ (1 + 3 \sin^2 \theta - 3 \sin^2 \chi)  + 3 \left( 1 - \sin^2 \theta - \sin^2 \chi \right) \gamma^5 \right] \, ,
\\[2mm]
g_{12}^{Z'}  &= g_{21}^{Z'} = \frac{3}{8} \, g_{_B}    \left[ \sin 2 \theta - \sin 2 \chi - (\sin 2 \theta + \sin 2 \chi ) \, \gamma^5 \right] \, ,
\\[2mm]
g_{22}^{Z'}  &= \frac{g_{_B}}{4} \left[ (1 - 3 \sin^2 \theta + 3 \sin^2 \chi)  - 3 \left( 1 - \sin^2 \theta - \sin^2 \chi \right) \gamma^5 \right] \, 
\label{eq:ZZ'_couplings-Zp}
\end{split}
\eear
for the $Z'$ boson, and 
\bear
\begin{split}
g_{11}^Z &=  \frac{g}{4 c_W}  \left[ \left( 4 s_W^2 - \cos^2 \theta - \cos^2 \chi \right)  - \left( \sin^2 \theta - \sin^2 \chi \right) \gamma^5 \right] \, ,
\\[2mm]
g_{12}^Z &= g_{21}^Z = \frac{g}{8 c_W} \left[ \sin 2 \theta + \sin 2 \chi + (-\sin 2 \theta + \sin 2 \chi) \, \gamma^5 \right] \, ,
\\[2mm]
g_{22}^Z &= \frac{g}{4 c_W}  \left[ \left( 4 s_W^2 - \sin^2 \theta - \sin^2 \chi \right)  + \left( \sin^2 \theta - \sin^2 \chi \right) \gamma^5 \right] \, 
\label{eq:ZZ'_couplings}
\end{split}
\eear
for the $Z$ boson.
Again, for later convenience, the above $Z'$ and $Z$ couplings are written in terms of vector and axial-vector couplings, denoted by 
\be
g_{ab}^{X} = g_{ab}^{X (V)}  + g_{ab}^{X  (A)}  \gamma^5  ~~,
\label{eq:gVgA}
\ee
where $X$ here stands for $Z'$ or $Z$.
It is straightforward to check that these couplings
reduce to the expected diagonal couplings in the limit 
$\theta, \chi \rightarrow 0$.

For completeness, we also briefly list the $W$ interactions with the
charged and neutral anomalons, in the limit of vanishing neutral
anomalon mixing when $y_3$ and $y_4$ are zero.  The charged current
Lagrangian is
\begin{align}
\mathcal{L}_{\text{W}} &= W_\mu^+ \bar{N}_2 \gamma^\mu \left(
g_{2}^W E_2 + g_{1}^W E_1 \right) + \text{ h.c.} \ .
\end{align}
The corresponding anomalon couplings to the $W$ boson are
\begin{align}
    \begin{split}
        g_{1}^W &= \frac{g}{2 \sqrt{2}} \left[ -\left( \sin \theta + \sin \chi \right) + \left( \sin \theta - \sin \chi \right) \gamma^5 \right]\ , \\
        g_{2}^W &= \frac{g}{2 \sqrt{2}} \left[ \left( \cos \theta + \cos \chi \right) - \left( \cos \theta - \cos \chi \right) \gamma^5 \right] \ .
    \end{split}
    \label{eq:WW_couplings}
\end{align}

\subsection{Scalar production at hadron colliders}
\label{subsec:varphiproduction}

\begin{figure}[t!]
 \begin{center}
  \hspace{-0.2cm}
 \includegraphics[width=0.39\textwidth, angle=0]{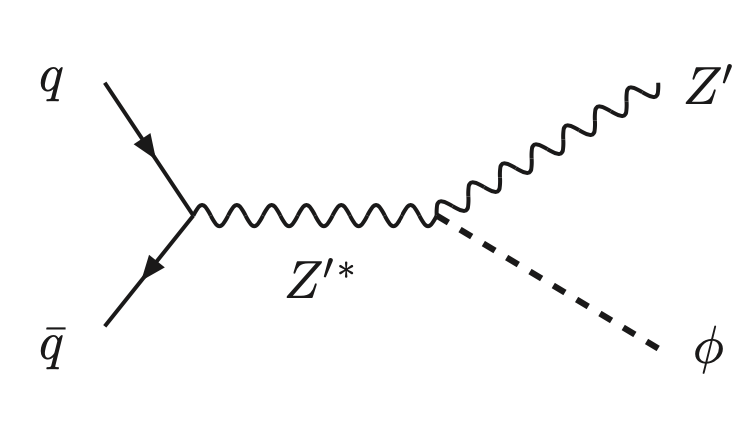}
 \hspace{2cm}
\includegraphics[width=0.37\textwidth, angle=0]{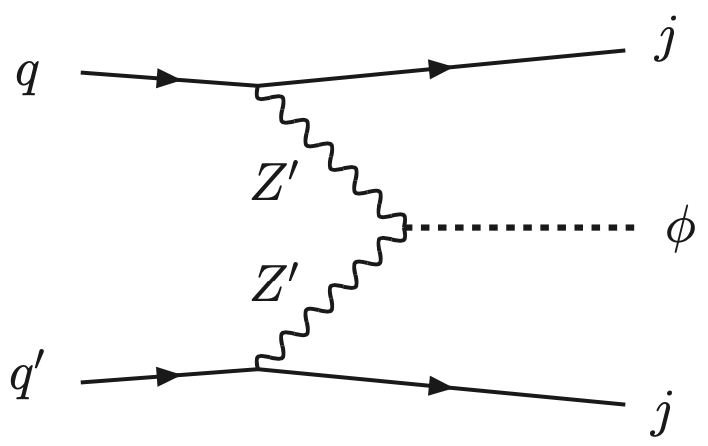}
 \caption{ Production processes of the $\phi$ scalar at hadron colliders:
   associated $\phi \, Z'$ production through an off-shell $Z'$ boson (left diagram), and $Z'$-fusion
   (right diagram).  }
\label{fig:phidiags}
\end{center}
\end{figure}

As the scalar associated with spontaneous breaking of a gauge 
symmetry, $\phi$ can be produced at the LHC via the
intrinsic and diagnostic processes of  $\phi$-strahlung from $Z'$ as
well as $Z'$ fusion, with Feynman diagrams shown in
Figure~\ref{fig:phidiags}.  Another set of scalar production modes
arise from anomalon pair production from electroweak interactions followed by
decays involving $\phi$, or by radiation of a $\phi$; as their
cross sections are highly dependent on anomalon masses, we will ignore
such processes in what follows.  

\begin{figure}[t!]
 \begin{center}
\hspace*{-0.3cm} 
 \includegraphics[width=0.8\textwidth, angle=0]{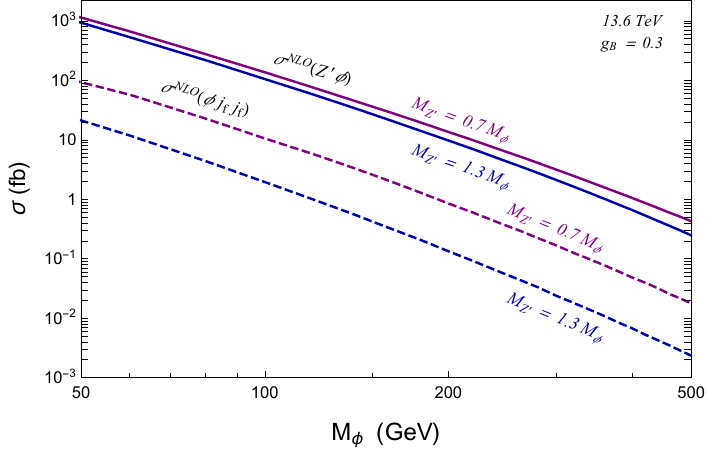}
 \caption{ Cross sections at the 13.6~TeV LHC for $\phi$ production in
   association with a $Z'$ (solid lines), and through $Z'$ fusion
   (dashed lines, labelled $\phi j_f j_f$, where $j_f$ is a forward
   jet), as a function of the $\phi$ mass.  The $Z' \phi$ and $\phi
   j_f j_f$ cross sections are computed at NLO in $\alpha_s$ for
   $M_{Z'}/M_{\phi}= 0.7$ (upper lines) and $1.3$ (lower lines), for a
   $Z'$ gauge coupling $g_{_B} = 0.3$; for fixed $M_{Z'}/M_{\phi}$, the
   cross section scales as $g_{_B}^4$ for $Z' \phi$ production, and as
   $g_{_B}^6$ for $Z'$ fusion.}
\label{fig:phixsecs_nomixing}
\end{center}
\end{figure}

We show the cross sections for $p p \to Z' \, \phi$ and $p p \to \phi
jj$ via $Z'$-fusion in Figure~\ref{fig:phixsecs_nomixing} for $g_{_B} =
0.3$ and two choices of the mass ratio, $M_{Z'} / M_{\phi} = 0.7$ or
$1.3$.  Notably, with $g_{_B} = 0.3$ at the 13.6~TeV LHC, the $Z' \, \phi$
cross section ranges from 1~fb to 1000~fb for $M_{\phi}$ ranging from
500~GeV to 50~GeV, while the $Z'$-fusion rate is roughly one or two orders of
magnitude smaller.  Given that $M_{Z'}$ is held fixed, the cross
section for $Z' \, \phi$ production scales as $g_{_B}^4$, while the 
$Z'$-fusion rate scales as $g_{_B}^6$.  

These rates were calculated using
MadGraph\_aMC@NLO~\cite{Alwall:2014hca} with model files generated by the FeynRules package~\cite{Alloul:2013bka}. The computation incorporates next-to-leading order (NLO) QCD corrections via FeynArts~\cite{Hahn:2000kx}, and uses the default PDF set employed  in Madgraph, 
NNPDF23\_nlo\_as\_0119\_qed  \cite{Ball:2013hta}. 

We remark that these production processes are intrinsic to the spontaneous breaking by $\Phi$ of any new $U(1)$ gauge symmetry,  and thus apply qualitatively to many $Z'$ models. The $\phi-Z'-Z'$ vertex evident in both processes is emblematic of the Higgs mechanism. 
Furthermore, both  cross sections  are independent of the anomalon masses or charges.
Nevertheless, the rates obtained for $U(1)_B$ cannot be directly used to compute the rates for different 
$Z'$ couplings to the up and down quarks (initial states with heavier quarks have negligible contributions). The reason is that  the ratio of the $u$ and $d$ PDFs depends on the energy scale of the process, which in turn depends on $M_{Z'}$ and $M_{\phi}$. 
  
 We now study the corresponding chiral symmetry breaking phenomenology for $\phi$ decays, assuming negligible mixing between the $\phi$ scalar and the SM Higgs boson. 
The situation when $\phi$ and $h^0_{\rm SM}$ have sizable mixing is analyzed in Section~\ref{sec:phiHiggs}.

\subsection{Branching fractions of the scalar $\phi$}
\label{subsec:varphidecays}

The $\phi$ scalar, as the radial mode of the scalar field that
spontaneously breaks a chiral gauge symmetry, exhibits a few classes
of possible decays.  

First, the $\phi$ scalar is expected to dominantly decay to a pair of $Z'$ bosons, with one of them or even both being off-shell for $M_{\phi} < 2 M_{Z'}$.  The subsequent tree-level decays of $Z' \to jj$ give
difficult but interesting 4-jet final states, where the 4-jet
invariant mass peaks at $M_\phi$, and include one or two dijet 
resonances peaked at $M_{Z'}$ when the corresponding $Z'$ is on-shell.
The partial width for $\phi$ decaying into a pair of $Z'$ bosons is given by (see~\cite{Djouadi:2005gi})
\begin{align}
\Gamma (\phi \to Z' Z'{}^{(*)}) &=
\left\{
\begin{array}{lc}
\dfrac{9 g_{_B}^2 }{128 \pi }  \, M_\phi \, R_2  \! \left( M_{Z'}^2/M_\phi^2 \right) 
  &  \hspace*{-0.2cm}  , \;\;   {\rm for}  \;\;  M_\phi > 2 M_{Z'} 
\\[0.5cm]
\dfrac{5 \, g_{_B}^4 }{ 1024 \pi^3}  \, M_\phi \,
R_T \! \left( M_{Z'}^2/M_\phi^2 \right)  & 
 , \;\;  {\rm for}  \;\;  2 M_{Z'} > M_\phi > M_{Z'}  ~~, 
\end{array}
\right. 
\label{eq:phi_ZpZp} 
\end{align}
where the second line is the 3-body decay $\phi \to Z' Z'{}^* \to Z'
q\bar q \, $ summed over 5 SM quark flavors; following
Ref.~\cite{Djouadi:2005gi}, we defined the functions
\be
R_2(x) = \left(12 x  - 4 + \frac{1}{x} \right) \sqrt{1 - 4x} \ ,
\ee
\be
R_T(x) = 3 \, \frac{1 - 8x + 20 x^2}{\sqrt{4x -1}} \arccos \! \left(
\frac{3x - 1}{2 x^{3/2}} \right) - \frac{1-x}{2x} (2 - 13 x + 47 x^2) 
- \frac{3}{2} (1 - 6 x + 4 x^2) \ln x ~.
\ee

As a second possibility, the $\phi$ scalar could in principle decay to pairs of
anomalons via the Yukawa couplings in Eq.~(\ref{eq:Ycouplings}), but here we
assume the anomalons are heavier than $M_\phi/2$, and thus
this decay channel is kinematically closed.
Another decay process is $\phi \to b\bar b$, which occurs at one loop, with two $Z'$ bosons running in the loop together with a $b$ quark. The partial width of this decay mode is suppressed not only by the loop factor, but also by $(m_b/M_{Z'})^2$. A similar $\phi \to t\bar t$ mode would occur through a  less suppressed loop, but it is kinematically closed for the range of $\phi$ masses considered here. 

More relevant are loop-induced decays of $\phi$ to pairs of gauge
bosons through loops of anomalons.  In particular, the anomalons give
non-decoupling contributions to loop-induced decays $\phi \to \gamma
\gamma$, $Z' \gamma$, $Z \gamma$, $Z' Z$, $ZZ$ and $W^+ W^-$,
where the last five modes can have additional phase-space suppression
factors.  Here, the nondecoupling of anomalons precisely follows the
Higgs low-energy theorem from SM Higgs
phenomenology~\cite{Shifman:1979eb, Kumar:2012ww}.  We also remark
that the $\phi \to Z' Z$ diagram explicitly reintroduces $Z'$-$Z$ mixing
at the anomalon mass scale.  Separately, for finite $\theta$ and
$\chi$, the anomalons introduce $\phi-h_{\text{SM}}^0$ mixing at one loop.
Nonetheless, vanishing $\phi-h_{\text{SM}}^0$ mixing at one-loop can still be
ensured by an appropriate choice of the Higgs portal coupling, hence
we neglect these mixing diagrams and the anomalon contributions to the
$W$ and $Z$ masses in this section.  We explicitly study the effect of
finite $\phi-h_{\text{SM}}^0$ mixing in the next section.

\begin{figure}[htb!]
\centering
\begin{subfigure}{0.4\textwidth}
\begin{tikzpicture}
\begin{feynman}
        \vertex (a) {\(\phi\)};
        \vertex [right=1.5cm of a]  (a1);
        \vertex [above right =2.0cm of a1] (a2a);
        \vertex [below right =2.0cm of a1] (a2b);
        \vertex [right=1.5cm of a2a] (b) {$V_1^\mu$};
        \vertex [right=1.5cm of a2b] (c) {$V_2^\nu$};
        \vertex [below right=1.5cm of c] (d);
        \vertex [above right=1.5cm of c] (e);
        \vertex[above = 0.03cm of a2a] (mu) {}; 
        \vertex[below = 0.03cm of a2b] (mu) {}; 
        
        \diagram* {
            (a) -- [scalar, momentum = \(P\)] (a1),
            (a1) -- [fermion, momentum = \(k+p_1\), edge label' = \(b\) ] (a2a),
            (a2a) -- [fermion, momentum =  \(k\), edge label' = \(a\) ] (a2b),
            (a2b) -- [fermion, momentum = \(k-p_2\), edge label' = \(c\) ] (a1),
            (a2a) -- [photon, momentum = \(p_1\)] (b),
            (a2b) -- [photon, momentum'=\(p_2\)] (c) 
        };
    \end{feynman}
\end{tikzpicture}
\caption{On-Shell two-body decay}
\label{fig:phiVVdecays_2bdy}
\end{subfigure}
\begin{subfigure}{0.4\textwidth}
\begin{tikzpicture}
\begin{feynman}
        \vertex (a) {\(\phi\)};
        \vertex [right=1.5cm of a]  (a1);
        \vertex [above right =2.0cm of a1] (a2a);
        \vertex [below right =2.0cm of a1] (a2b);
        \vertex [right=1.5cm of a2a] (b) {$V_1^\mu$};
        \vertex [right=1.5cm of a2b] (c);
        \vertex [below right=1.5cm of c] (d);
        \vertex [above right=1.5cm of c] (e);
        \vertex[above = 0.03cm of a2a] (mu) {};
        \vertex[below = 0.03cm of a2b] (mu) {};
        \vertex[above right = 0.2cm and 1.2cm of c] (l1) {$f$};
		\vertex[below right = 0.2cm and 1.2cm of c] (l2) {$\bar{f}$};
		\vertex[above right=0.6cm of a2b, yshift=-0.4cm] (f) {$V_2$};
        
        \diagram* {
            (a) -- [scalar, momentum = \(P\)] (a1),
            (a1) -- [fermion, momentum = \(k+p_1\), edge label' = \(b\) ] (a2a),
            (a2a) -- [fermion, momentum =  \(k\), edge label' = \(a\) ] (a2b),
            (a2b) -- [fermion, momentum = \(k-p_2\), edge label' = \(c\) ] (a1),
            (a2a) -- [photon, momentum = \(p_1\)] (b),
            (a2b) -- [photon, momentum'=\(p_2\)] (c), 
            (c) -- [fermion, momentum = \(q_2\) ] (l1),
            (c) -- [anti fermion, momentum' = \(q_3\) ] (l2)
        };
    \end{feynman}
\end{tikzpicture}
\caption{Decay to three on-shell SM states}
\label{fig:phiVVdecays_3bdy}
\end{subfigure}
\caption{Feynman diagrams for $\phi$ decay to (a) generic vector boson
  final states $V_1$, $V_2$ and (b) $V_1 f \bar{f}$, where the intermediate fermions labeled
  $a$, $b$, and $c$ denote all possible combinations of anomalons. The diagrams with
  reversed fermion flow are implicitly included.}
  \label{fig:phiVVdecays}
\end{figure}
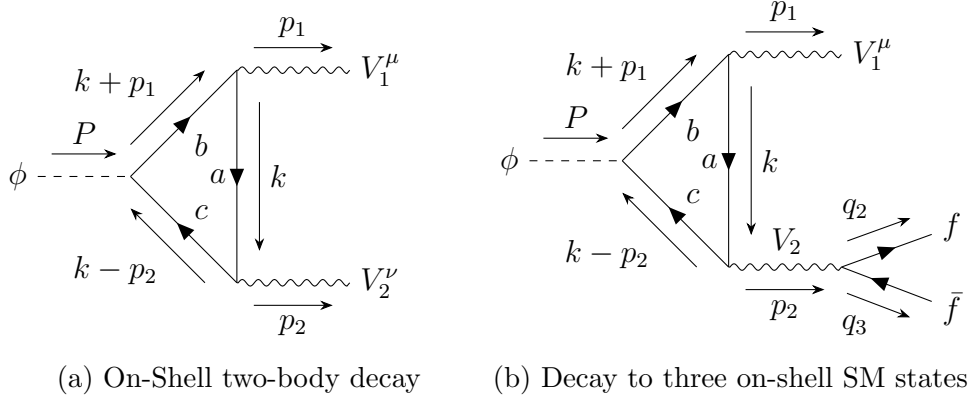

The 1-loop decay of $\phi$ into photons gets contributions from the
$E_1$ and $E_2$ charged anomalons, giving
\begin{flalign} 
    \Gamma (\phi \to \gamma \gamma) &=
        \frac{\alpha^2 M_\phi^3}{256 \pi^3} \left|
            \sum_{a = 1}^2 \frac{y_{aa}^{\phi (S)}}{M_a} F_1 
            \!  \left( \frac{M_\phi^2}{4 M_a^2} \right) 
        \right|^2   ~~,
\end{flalign}
where $y_{aa}^{\phi (S)}$ are given in Eqs.~(\ref{eq:ySyP}) and (\ref{eq:couplings}). The above loop function is~\cite{Djouadi:2005gi}
\begin{flalign}
F_1(x) &= \frac{2}{x^2} \left[ x + (x-1) Z(x) \rule{0mm}{3.9mm} \right]   ~~,
\label{eq:F1} 
\end{flalign}
with
\begin{flalign}
Z(x) &= 
\left\{
\begin{array}{cc}
\dfrac{-1}{4} \left[ 2\ln \! \left(\sqrt{x}  + \sqrt{x-1}\,  \rule{0mm}{3.87mm}  \right) - i \pi  \rule{0mm}{3.9mm} \right]^2
&  \hspace*{.5cm} , \;\;  {\rm for}  \;\;  x > 1 \\ [2mm]
\arcsin^2 (\sqrt{x}) &  \hspace*{.5cm} , \;\;  {\rm for}  \;\;  x \leq 1 \\ [2mm]
\end{array} \right.
\ . 
\label{eq:Zx}
\end{flalign}
There are also intrinsic 1-loop decays into $Z' \gamma$ and $Z \gamma$, with decay widths given by
\be
\Gamma(\phi \to X \gamma) = \frac{\alpha M_\phi^3 (1-M_{X}^2/M_\phi^2)^3}{32 \pi^4} 
\left| \sum_{a,b = 1}^2 \frac{c^{(X)}_{ab}}{M_a}   F^{X \gamma} (M_\phi, M_{X}, M_a, M_b) \right|^2 ~~,
\ee
where $X = \left\{ Z', Z \right\}$, and the coefficients $c^{(X)}_{ab}$ are defined as
\be
c^{(X)}_{ab} = y_{ab}^{\phi (S)} g_{ab}^{X (V)} - y_{ab}^{\phi (P)} g_{ab}^{X (A)}   ~~.
\label{eq:cab}
\ee
Here $y_{ab}^{\phi}$ and $g_{ab}^{X}$ are the Yukawa and gauge
couplings given in Eqs.~(\ref{eq:ySyP}), (\ref{eq:gVgA}),  
respectively, and the loop function is
\bear 
&&  \hspace*{-0.7cm}
F^{X \gamma} \!(M_\phi, M_{X}, M_a, M_b) 
 = \!\frac{ 2 M_a^2}{M_\phi^2 \! -\! M_{\tilde{Z}}^2} \! \left[  \frac{B(M_\phi^2, M_a, M_b)\! - \! B(M_{X}^2, M_a, M_b)}{1- M_\phi^2/M_{X}^2} 
  	- \frac{M_a^2 \! - \! M_b^2}{M_\phi^2} \ln\!\left( \! \frac{M_a}{M_b} \!\right) \right.
 \nonumber \\
&& \left.   \hspace*{-0.3cm}  + 
		\left( \frac{M_\phi^2 \! - \! M_{X}^2}{2} -  \! M_a^2 \! \right) C_0(0, M_\phi^2, M_{X}^2, M_a, M_a, M_b) 
-  M_b^2 \, C_0(0, M_\phi^2, M_{X}^2, M_b, M_b, M_a) -1
\right]  \nonumber \\
&&
\label{eq:FVgamma_general}
\eear
with $B$, $C_0$ being Passarino-Veltman loop
functions~\cite{Passarino:1978jh} following the Package-X
convention~\cite{Patel:2015tea}.  
For $M_a = M_b$,
$F^{X \gamma} (M_\phi, M_{X}, M_a, M_a) =  F_2
\left( \frac{4 M_a^2}{M_{\phi}^2}, \frac{4 M_a^2}{M_{X}^2} \right)$,
where the $F_2$ loop function is adapted from~\cite{Djouadi:2005gi}:
\begin{flalign}
 \hspace*{-.3cm}
F_2 (x, y) =  \frac{ x y}{2(x-y)}  \Bigg[ 1 &+  \left(1+\frac{x y }{x-y} \right) \!\left(Z( x^{-1} ) - Z( y^{-1} )  \rule{0mm}{3.9mm} \right) 
+ 2x \, \frac{ g( x^{-1} ) - g( y^{-1} ) }{x-y}\Bigg] \ ,
\label{eq:F2}
\end{flalign}
with 
\begin{flalign}
g(x) &=
\left\{
\begin{array}{cc}
\sqrt{1/x - 1} \arcsin (\sqrt{x}\, ) &  \hspace*{.5cm} , \;\;  {\rm for}  \;\;   x \leq 1 \\ [2mm]
\sqrt{1 - 1/x}  \left[  \ln \! \left(\sqrt{x} + \sqrt{x-1} \,  \rule{0mm}{3.87mm} \right) - i \pi/2    \rule{0mm}{3.9mm}  \right] & 
 \hspace*{.5cm} , \;\;  {\rm for}  \;\;   x > 1 \\
\end{array}   ~~.
\right. 
\label{eq:gx}
\end{flalign}
Lastly, $\phi$ also decays into pairs of massive gauge bosons at
one-loop.  Since the charged anomalons have both $U(1)_B$ and EW vev
contributions to their masses, their couplings to $\phi$, $Z$, and
$Z'$ are generally flavor-violating, as shown in
Sec.~\ref{subsec:phi}.  Hence, we present the corresponding decay
width calculations for $\phi \to Z' Z$, $ZZ$, and $W^+ W^-$ with the
reference matrix elements shown in Fig.~\ref{fig:phiVVdecays_2bdy}
for $\phi$ decaying to two on-shell massive vectors and
Fig.~\ref{fig:phiVVdecays_3bdy} for $\phi$ decaying to a three-body
final state of SM particles.

The matrix elements for Fig.~\ref{fig:phiVVdecays} are then 
\begin{align}
    \begin{split}
        i \mathcal{M}_{abc} (\phi \to V_1 V_2) &=
        -\epsilon^*_\mu (p_1) \epsilon^*_\nu( p_2) \int \frac{d^4 k}{(2\pi)^4}
        \\
        & \mathrm{Tr} 
        \left[
            \frac{
                \slashed{k} + M_a}{k^2 + M_a^2}
            \gamma^\mu(i g_{ab}^{V_1})
            \frac{\slashed{k} + \slashed{p}_1 + M_b}{(k+p_1)^2 - M_b^2}
            (-i y_{bc}^\phi )
            \frac{\slashed{k} - \slashed{p}_2 + M_c}{(k-p_2)^2-M_c^2}
            \gamma^\nu (i g_{ca}^{V_2})
        \right]
    \end{split}
    \label{eq:matrixElement_2bdy}
\end{align}

and 
\begin{align}
    \begin{split}
        i \mathcal{M}_{abc} (\phi \to V_1 f \bar{f}) &=
        -\epsilon^*_\mu (p_1) \left( \frac{- i g_{\nu \alpha}}{(q_2 + q_3)^2 - M_{V_2}^2} \bar{u}(q_2) \gamma^\alpha \left( i g_{f\bar{f}}^{V_2}  \right) v(q_3) \right) \int \frac{d^4 k}{(2\pi)^4}
        \\
        & \mathrm{Tr} 
        \left[
            \frac{
                \slashed{k} + M_a}{k^2 + M_a^2}
            \gamma^\mu(i g_{ab}^{V_1})
            \frac{\slashed{k} + \slashed{p}_1 + M_b}{(k+p_1)^2 - M_b^2}
            (-i y_{bc}^\phi )
            \frac{\slashed{k} - \slashed{p}_2 + M_c}{(k-p_2)^2-M_c^2}
            \gamma^\nu (i g_{ca}^{V_2})
        \right] \ ,
    \end{split}
    \label{eq:matrixElement_3bdy}
\end{align}
where $V_1$ and $V_2$ refer to $Z'Z$, $ZZ$, and $W^+ W^-$, respectively.
For the three-body decay, we assume that the final state SM fermions
are massless, and the partial width coherently sums over all
accessible SM fermions with their respective gauge couplings to $W$ or
$Z$.  The total matrix elements are readily computed from the above
generic matrix elements, taking into account the diagrams with reversed fermion flow
and the flavor-conserving and
flavor-violating couplings shown in Sec.~\ref{sec:theory}.  For
completeness, we also show the corresponding phase space integrations
for the two-body $\phi \to V_1 V_2$ partial decay,
\begin{align}
        \Gamma(\phi \to V_1 V_2)  =
        \frac{1}{16 \pi M_\phi^3}  \sqrt{\lambda(M_\phi^2, M_{V_1}^2, M_{V_2}^2)} \left| \mathcal{M}(\phi \to V_1 V_2 ) \right|^2
\end{align}
with $\lambda$ denoting the K\"{a}ll\'{e}n-$\lambda$ function, and the $\phi \to V f\bar{f}$ partial decay, 
\begin{align}
    \Gamma(\phi \to V_1 f \bar{f}) = \frac{1}{(2 \pi)^3} \frac{1}{32 M_\phi^3} 
    \int_{M_{V_1}^2}^{M_\phi^2} d M_{12}^2 \int_{0}^{4 E_f E_{\bar{f}} } d M_{23}^2
    \left| \mathcal{M}(\phi \to V_1 f \bar{f} ) \right|^2
    \label{eq:decay_width_PDG}
\end{align}
where we used
\begin{align}
    E_f E_{\bar{f}} \approx \frac{1}{4 M_{12}^2} \left( M_{12}^2 - M_{V_1}^2 \right) \left( M_\phi^2 - M_{12}^2 \right) \ ,
\end{align}
and $M_{12}^2 = (p_1 + q_2)^2$ and $M_{23}^2 = (q_2 + q_3)^2$.  All of
these expressions are the dominant decay widths necessary to
understand the collider phenomenology of $\phi$ decay modes when
$\phi-h_{\text{SM}}^0$ mixing is absent but where anomalon mixing angles $\theta$
and $\chi$ are present.

\begin{figure}[t!]
 \begin{center}
 \hspace*{-0.3cm} 
  \includegraphics[width=0.8\textwidth, angle=0]{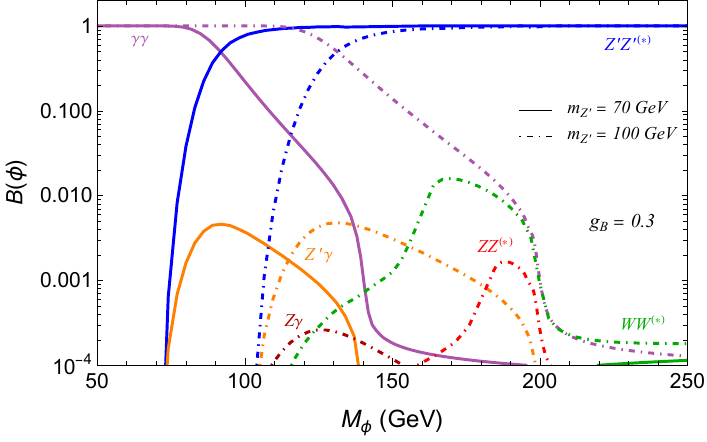}
   \caption{Branching fractions for the $U(1)_B$-breaking scalar
     $\phi$ as a function of its mass.  The parameters fixed here are
     $g_{_B} = 0.3$, $M_{Z'} = 70$~GeV (solid lines) or $M_{Z'} =
     100$~GeV (dashed lines), and negligible mixing with the SM Higgs
     boson.  For the loop decays, $\phi \to \gamma \gamma$, $Z'
     \gamma$, $Z \gamma$, $ZZ^{(*)}$, and $WW^{(*)}$, we set the lighter anomalon mass to 200~GeV and the heavier to 250~GeV. The
     anomalon mixing angles are fixed to $\theta = 0.3$ and $\chi =
     0.25$.}
\label{fig:brs_mphi_nomixing}
\end{center}
\end{figure}

The branching fractions of $\phi$ are shown in
Figure~\ref{fig:brs_mphi_nomixing} as a function of $\phi$ mass and
two choices of $M_{Z'} = 70$~GeV or $100$~GeV. For both choices of
$M_{Z'}$, we set one charged anomalon mass to $200$~GeV and the other to 
$250$~GeV, with anomalon mixing angles of $\theta = 0.3$ and $\chi =
0.25$, adjusting the Yukawa couplings accordingly. We remark that
the resulting Yukawa couplings are hierarchical with $y_1, y_2 \ll y_L,
y_E$ and thus the modification of $B(h \to \gamma \gamma)$ is less
than one percent~\cite{Michaels:2020fzj}. In addition, the anomalons can be
produced at the LHC via Drell-Yan processes, but their collider signatures
into charged pions and missing transverse energy are difficult to detect~\cite{Dobrescu:2021vak}.

For the $\phi$ mass range
shown, the nominally leading decay, $\phi \to Z' Z'{}^{(*)}$, can be a
3-body process and becomes kinematically suppressed for lighter
$M_{\phi}$.  Hence, the loop-induced decay $\phi \to \gamma \gamma$
becomes comparable to the $Z' Z'{}^{(*)}$ decay and in fact dominates
for light $\phi$ masses.  The other loop-induced decays to $Z'
\gamma$, $W^+ W^-$, and $ZZ$ can also have percent-level
branching fractions, and their complementary collider signatures are
important to detect in order to extract the underlying pattern of
anomalon masses and couplings.  At a subleading rate, we can also expect the $Z \gamma$ decay.  Importantly, the $\phi \to Z f\bar{f}$ partial width is mediated by one-loop amplitudes $\phi \to Z Z'^{*}$ and $\phi \to Z Z^{*}$ and their interference, which is included in Figure~\ref{fig:brs_mphi_nomixing} (where the final state SM fermions are assumed massless).  We have verified that the interference with the loop-induced coupling $\phi \to Z \gamma$, with $\gamma$ off-shell, is negligible.  The branching fraction for $\phi \to Z'Z$ falls below $10^{-4}$ and is not shown.  We remark that the branching ratio of $\phi \to WW^{(*)}$ and $\phi \to ZZ^{(*)}$ compared to $\phi \to Z'Z'^*$ reflects the mass hierarchy of the electroweak gauge boson masses compared to the $Z'$ mass.

This pattern of decays from on-shell $Z' Z'$ for
large $\phi$ masses to $\gamma \gamma$ for small $\phi$ masses is
characteristic of scalars spontaneously breaking chiral gauge
symmetries, and is reminiscent of SM Higgs phenomenology.  In
particular, for fixed gauge and chiral fermion masses, the heavy
scalar limit necessarily corresponds to an enhanced coupling to
longitudinal gauge bosons and the on-shell decay to two vector bosons
saturates the decay width of the scalar.  At the opposite end, the
light scalar limit has a total decay width dominated by the
nondecoupling loop effects of chiral fermions, which mediate decays
to any and all kinematically allowed spectator bosons.

\subsection{LHC signals of  the $\phi$ scalar}

Having presented the LHC production modes and decay widths for $\phi$,
we now discuss the collider phenomenology of the $\phi$ scalar
responsible for $U(1)_B$ breaking.  Since the cross sections for
$\phi$ production (see Figure~\ref{fig:phixsecs_nomixing}) fall
rapidly with increasing $\phi$ and $Z'$ masses, we focus on mass
ranges below a few hundred GeV.  To analyze the LHC signals, we
consider three intervals for the $M_\phi/M_{Z'}$ ratio.

\begin{figure}[t!]
 \begin{center}
 \includegraphics[width=0.4\textwidth, angle=0]{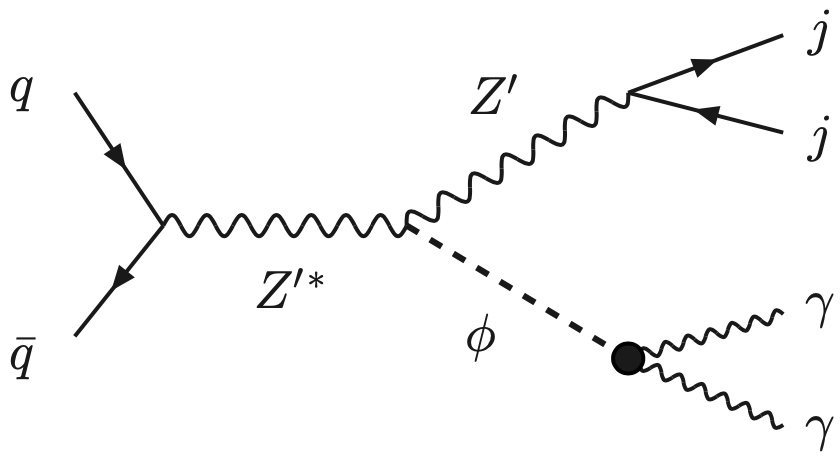}
 \hspace*{1.6cm} 
\includegraphics[width=0.42\textwidth, angle=0]{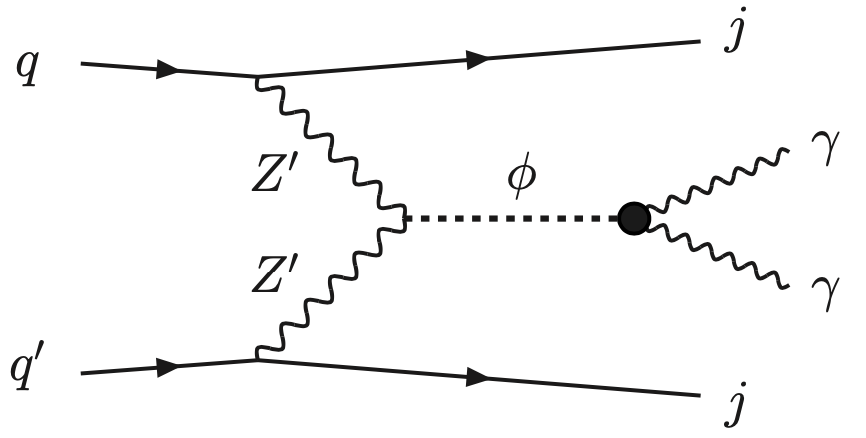} 
\vspace*{-0.1cm} 
 \caption{LHC signals of $\phi$ in the low-$M_\phi/M_{Z'}$ range,
   $M_\phi \lesssim 1.5 M_{Z'}$: $(\gamma\gamma)(jj)$ (left diagram)
   and $(\gamma\gamma)j_f j_f$ (right diagram). The $(jj)$ notation
   refers here to a dijet resonance at $M_{Z'}$, while $j_f$ is a
   forward jet, as typically produced in $Z'$ fusion.
}
\label{fig:lowphidiags}
\end{center}
\begin{center}
 \includegraphics[width=0.468\textwidth, angle=0]{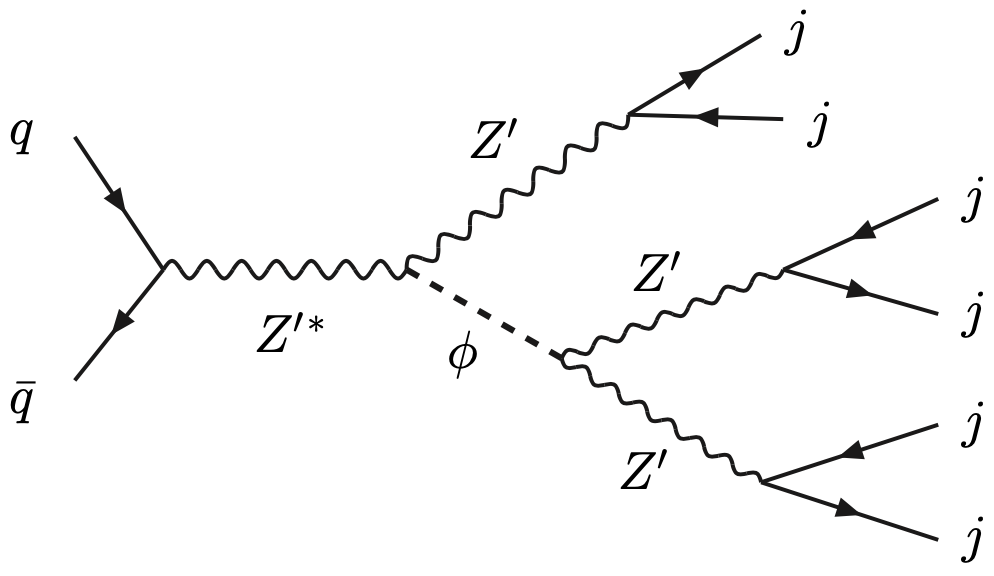}
 \hspace*{0.4cm} 
\includegraphics[width=0.488\textwidth, angle=0]{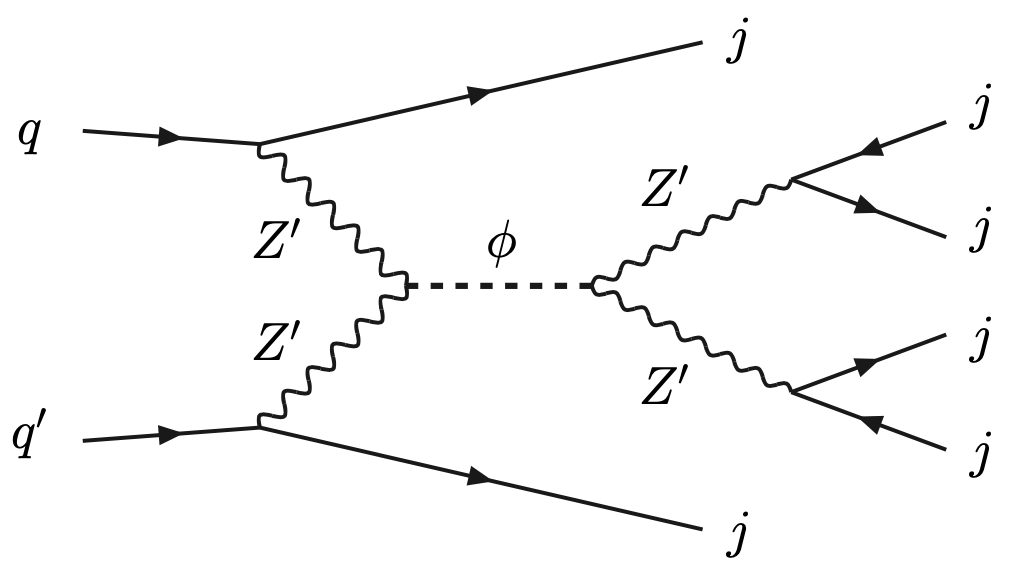}
\vspace*{-0.1cm} 
 \caption{LHC signals of $\phi$ in the high-$M_\phi/M_{Z'}$ range,
   $M_\phi > 2 M_{Z'}$: $3(jj)$ (left diagram) and $(jj)(jj)j_f j_f$
   (right diagram).
 }
\label{fig:highphidiags}
\end{center}
\end{figure}

\subsubsection{Low $M_\phi/M_{Z'}$}

For the ``low-$M_\phi/M_{Z'}$ range", $M_\phi \lesssim 1.5 M_{Z'}$,
the main decay mode of $\phi$, with a branching fraction above $60\%$,
is $\phi \to \gamma\gamma$.  In this case associated $\phi$ production
leads to a signal that involves a $\gamma\gamma$ resonance at $M_\phi$
plus a dijet resonance at $M_{Z'}$, as shown in the left panel of
Figure~\ref{fig:lowphidiags}.  The $\phi$ production via $Z'$ fusion
also leads to a signal (see right panel of
Figure~\ref{fig:lowphidiags}) with a $\gamma\gamma$ resonance at
$M_\phi$, but the additional two jets do not form a resonance, and are
mostly in the forward region.  We will refer to these two signals as
$(\gamma\gamma)(jj)$ and $(\gamma\gamma)j_f j_f$, respectively, where
$(jj)$ denotes a dijet resonance at $ M_{Z'}$, while $j_f$ denotes a
forward jet as typically produced in $Z'$ fusion.

To illustrate the LHC capabilities in the low-$M_\phi/M_{Z'}$ range,
we consider two points in the parameter space: $g_{_B} =
0.3$, $M_\phi = 100$ GeV, and $M_{Z'} = 70 $ GeV or $M_{Z'} = 130 $
GeV.  From Figures \ref{fig:phixsecs_nomixing} and
\ref{fig:brs_mphi_nomixing} follows that for $M_{Z'} = 70$ GeV the
cross section at $\sqrt{s} =13.6\; {\rm TeV}$ times branching fraction
for $\phi \to \gamma\gamma$ is $\sigma (pp \to Z' \phi)  B
(\phi \to \gamma\gamma) \approx 130 $ fb.  Thus, about $4 \times 10^4$
$(\gamma\gamma)(jj)$ events will be produced in Run 3 of the LHC,
while the number of $(\gamma\gamma) j_f j_f$ will be smaller by a
factor of approximately 10.  The latter signal has only been studied
by CMS and ATLAS in the context of the SM Higgs~\cite{CMS:2022wpo,
  ATLAS:2023tnc}, or when exotic final states such as dark photons are
combined with forward jet tags~\cite{CMS:2020krr}.  Our analysis
demonstrates that $Z'$-fusion provides a parametrically new analysis
channel for nonstandard scalar production at the LHC.

Requiring two photons whose invariant mass distribution forms a peak
reduces considerably the background, which is due to real photon
emission but also due to QCD with jets misidentified as photons.  The
additional two jets, which either form a resonance at $M_{Z'}$ or are
forward, may be used to further reduce the background.

\subsubsection{High $M_\phi/M_{Z'}$}

For the ``high-$M_\phi/M_{Z'}$ range" $M_\phi > 2 M_{Z'}$, the
dominant decay mode is $\phi \to Z' Z'$, with a branching fraction
above $99\%$.  Associated $\phi$ production leads to a $3 Z'$ signal,
{\it i.e.}, six jets arranged in three dijet resonances, each of them
at $M_{Z'}$, as shown in the left panel of Figure~\ref{fig:highphidiags}.  
We refer to this as the $3(jj)$ signal, where $(jj)$ is
a dijet resonance at $M_{Z'}$, but note that two of the dijets also
form a 4-jet resonance at $M_\phi$.  Production of $\phi$ via $Z'$
fusion gives a $Z' Z' j_f j_f$ signal, {\it i.e.}, six jets arranged
in two dijet resonances at $M_{Z'}$ plus two forward jets, with the
two dijets forming a 4-jet resonance at $M_\phi$, as shown in the right 
panel of Figure~\ref{fig:highphidiags}.  We refer to this as
the $(jj)(jj)j_f j_f$ signal.

These signals suffer from large QCD backgrounds.  These may be
mitigated by requiring that one of the $Z'$ decays into $b\bar b$. As
the $Z' \to b\bar b$ branching fraction is 20\%, the $(jj)(jj)(b\bar
b) $ signal (where $j$ may or may not be a $b$ jet) has a combined
branching fraction of 49\%, while $(jj)(b\bar b) j_f j_f$
signal has a combined branching fraction of 36\%.  It is also possible
to consider boosted topologies to reduce the background, again at the
expense of a lower signal.  In the case of associated $\phi$
production, if the $\phi \to (jj)(jj)$ system is boosted due to a high
$p_T$ of the $Z' \to (jj)$ system, then the final state would involve
two wide jets, one with 4-prong substructure, and one with a 2-prong
substructure.  Another possibility is to require an initial jet
radiated with high $p_T$, so that either $\phi$ or $Z'$ or both are
boosted.

\subsubsection{Intermediate $M_\phi/M_{Z'}$}

In the intermediate-$M_\phi/M_{Z'}$ range $1.5 \lesssim M_\phi/M_{Z'}
< 2 $, the competition between the $\phi \to \gamma\gamma$ and $\phi
\to Z' Z^{\prime \, *}_B \to Z' jj$ branching fractions makes it
useful to search for both the lower-background signals involving a
pair of photons and the $6j$ signals. The latter are labelled
$(jj)(jj)jj$ and $(jj)(jj)j_f j_f$, while the former are labelled the
same way as in the low-mass case.  For the $6j$ signals, the
background reduction methods discussed for the high-mass range also
apply here, with the only difference that only two of the four narrow
jets inside the $\phi$ wide jet form a resonance at $M_{Z'}$.  On the
other hand, the diphoton final state follows the same strategy as the
low $M_\phi / M_{Z'}$ case.  We remark that the 1-loop decays to $ZZ$ 
and $WW$ could potentially be seen in Higgs observables, although the expected
rates imply that these effects would not serve as discovery channels.

To close this section, we remark that our focus on $\phi$ and $Z'$
masses below a few hundred GeV precludes $Z' \to t\bar{t}$ decays. 
However, the case where $M_{Z'} \gtrsim 400 $ GeV is also interesting (albeit the branching fraction for $Z' \to t\bar{t}$ remains below 1/6). In particular, for  
$M_\phi < M_{Z'}$ that case leads to the $(\gamma \gamma)(t\bar{t})$ signature.

Having analyzed the situation where $\phi$ has negligible mixing with
the SM Higgs boson, we now consider the collider phenomenology when
mixing is included.

\section{Effects of Higgs mixing} 
\setcounter{equation}{0}
\label{sec:phiHiggs}

In this section, we consider the phenomenology of the new scalar in
the presence of the Higgs portal coupling $|\Phi|^2 H^\dagger H $. The
mass mixing between the $\phi$ scalar and the SM Higgs boson $h^0_{\rm
  SM}$ leads to two mass eigenstates: the new physical scalar
$\varphi$ and $h^0$, which is identified with the discovered Higgs
boson with a mass near $125$~GeV.  Aside from mass mixing, the Higgs
portal coupling leads to exotic decays such as $h^0 \to \varphi
\varphi$ or $\varphi \to h^0 h^0$, depending on the scalar masses.

We begin with the full scalar potential,
\begin{align}
V(\Phi, H) &=  \lambda_\Phi \left( |\Phi|^2 - \frac{v^{\prime 2}_0}{2} \right)^2 
+ \lambda_H \left( H^\dagger H  - \frac{v^{2}_0}{2} \right)^2  + 2 \lambda_p |\Phi|^2 H^\dagger H \ ,
\end{align}
where the mass parameters $v_0$ and $v'_0$ are real and can be taken
positive, the dimensionless quartic couplings satisfy $0 <
\lambda_\Phi , \lambda_H \lesssim 1$, and the portal coupling
$\lambda_p$ is a real dimensionless parameter with $|\lambda_p| \ll
1$.  Minimizing the potential gives the vacuum expectation values
(vevs) $\langle \Phi \rangle = v' / \sqrt{2}$ and $\langle H \rangle =
(0, v_h )^\top / \sqrt{2}$, with
\begin{eqnarray}
&& v^{\prime 2} 
= \lambda_H \,  \frac{   \lambda_\Phi \, v^{\prime 2}_0  - \lambda_p  \,  v_0^2 }{  \lambda_H   \,  \lambda_\Phi - \lambda_p^2 }  ~~,
\nonumber \\ [-2mm]
&&   \label{eq:vevs}
\\ [-2mm]
&& v_h ^2 = \lambda_\Phi   \,  \frac{  \lambda_H   \,   v_0^2  -  \lambda_p  \, v^{\prime 2}_0   }{  \lambda_H   \, \lambda_\Phi - \lambda_p^2  }  ~~ .
\nonumber
\end{eqnarray}
Expanding $V(\Phi, H) $ around the vevs gives the scalar
mass terms 
\begin{align}
- \frac{1}{2} \left( \phi \, , \,  h_{\text{SM}}^0  \rule{0mm}{3.9mm} \right)
\left(
\begin{array}{cc}
2 \lambda_{\Phi} \, v'^2 & 2 \lambda_p  \, v_h v' \\
2 \lambda_p \, v_h  v' & 2 \lambda_H  \, v_h^2 
\end{array}
\right)
\left( \begin{array}{c}
 \phi \\ h_{\text{SM}}^0 
\end{array} \right) \ ,
\end{align}
leading to a scalar mixing angle $\alpha_h$ which satisfies
\begin{align}
\tan (2 \alpha_h ) = \dfrac{ 2 \lambda_p \, v_h v'}{\lambda_\Phi \, v'^2 - \lambda_H \, v_h^2} \ ,
\end{align}
and defines the scalar mass eigenstates as
\begin{align}
\left( \begin{array}{c}
\varphi \\  h^0
\end{array} \right)
=
\left( \begin{array}{cc} 
\cos \alpha_h & \sin \alpha_h \\
-\sin \alpha_h & \cos \alpha_h
\end{array}
\right)
\left(
\begin{array}{c}
\phi \\ h_{\text{SM}}^0
\end{array}
\right) \ .
\end{align}

The properties of the physical particle $h^0$ are modified from the SM
predictions by the above mixing.  Given the good agreement of the SM
predictions with the increasingly precise measurements of the Higgs
production cross sections and branching fractions performed by the
ATLAS and CMS collaborations, it must be that $|\sin \alpha_h|^2 \ll
1$.  Thus, we can expand in $\alpha_h$, which simplifies the relations
between physical observables and the vevs.  We obtain
\be
\sin \alpha_h \approx \dfrac{ \lambda_p \, v_h v'}{\lambda_\Phi \, v'^2 - \lambda_H \, v_h^2} ~~,
\ee
and the scalar physical masses:
\bear
&& M_\varphi  \simeq \sqrt{\lambda_\Phi} \,  v'   \left( 1 +  \dfrac{ 2 \lambda_p \, v_h}{ \lambda_\Phi v'} \, \sin \alpha_h
 \right) ~~,
\nonumber \\ [-2mm]
&&   \label{eq:Mvarphi}
\\ [-2mm]
&&  M_h \simeq  \sqrt{\lambda_H} \,   v_h \left( 1 -  \dfrac{ 2 \lambda_p \, v'}{ \lambda_H  v_h}\, \sin \alpha_h
\right) ~~.
\nonumber
\eear
Because the two scalar states are widely separated in mass, their
effects on observables are factorized on their distinct pole masses.
The complications arising when the mixing angle is large as well as
when the states are nearly degenerate were recently addressed
in~\cite{LoChiatto:2024guj}.

In the remainder of this Section, we analyze the effects of the
trilinear scalar interactions, and then the modifications of the Higgs
boson properties.

\subsection{Higgs decay $h^0 \rightarrow \varphi \varphi^{(*)}$ or 
resonant diHiggs process $\varphi \rightarrow h^0 h^0$}

The extended Higgs Lagrangian not only mixes $h_{\text{SM}}^0$ and $\phi$, it also
introduces three-point vertices including combinations of $h^0$ and
$\varphi$.  For $M_h > M_\varphi$, this results at leading order in
3-body decays of the SM-like Higgs into $\varphi$ and a pair of SM
particles, whose width is highly dependent on the accessible phase
space. For $M_h > 2 M_\varphi$, both scalars are on-shell which
increases the Higgs width by a tree-level decay into a $\varphi$ pair.
The $\varphi \varphi h^0$ and $\varphi h^0 h^0$ couplings in the Lagrangian can be written as
\begin{equation}
    - \mu_{\varphi \varphi h} 
    \, \varphi \, \varphi \, h^0
    - \mu_{\varphi h h} 
    \, \varphi \, h^0 \, h^0  ~~,
\end{equation}
where the trilinear couplings have
mass-dimension one.  Keeping in mind that there may be large
hierarchies between $ \lambda_p$, $\lambda_H$, $\lambda_\Phi$, and
also that the ratio $v_h/v' $ may be small, we expand in $\sin
\alpha_h$ and find
\begin{align}
\mu_{\varphi \varphi h } \simeq \lambda_p v_h - (3 \lambda_\Phi - 2 \lambda_p) v' \sin \alpha_h \ .
\end{align}
For reference, the coupling
\begin{align}
\mu_{\varphi h h} \simeq \lambda_p v' + (3 \lambda - 2 \lambda_p) v_h \sin \alpha_h \ ,
\end{align}
in the small angle approximation.

\begin{figure}[t]
\centering
    \includegraphics[width = 0.38\textwidth]{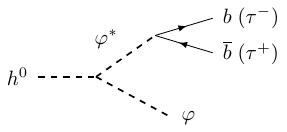}
    \caption{Three-body decay of the SM-like Higgs boson.}
    \label{fig:3:Body:Decay}
\end{figure}

For the mass interval $ M_h/2 < M_\varphi < M_h $, the SM-like Higgs
boson can have 3-body decays
\be
h^0 \to \varphi \, b \bar{b} \; {\rm or } \; \varphi \, \tau^+ \tau^- ~~.
\label{eq:3bodyprocess}
\ee 
Other tree-level 3-body decays are suppressed either by phase-space
({\it e.g.}, $h^0 \to \varphi \, W^{(*)} W^*$) or by smaller couplings
({\it e.g.}, $h^0 \to \varphi \, c \bar c$).  The one-loop 3-body
decay $h^0 \to \varphi \, gg$ is less interesting due to larger
backgrounds; its contribution to the Higgs width is only about 10\% of
the contributions from (\ref{eq:3bodyprocess}), and we will ignore it
in what follows.

\begin{figure}[t]
    \centering
    \includegraphics[width = 0.8\textwidth]{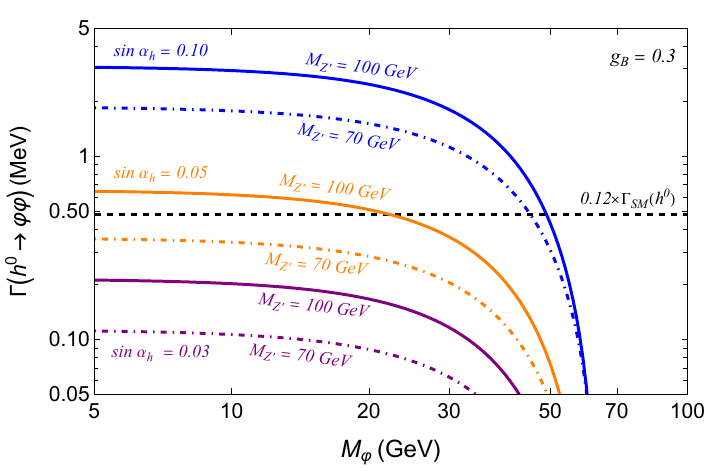}
    \caption{Decay width of the SM-like Higgs boson into $\varphi
      \varphi^{ (*) }$, including the 3-body decay through an
      off-shell $\varphi$ for $M_\varphi \lesssim M_h/2$.  The
      parameters fixed here are $M_{Z'} = 70$ GeV (solid) or 100 GeV
      (dot-dashed), $g_{_B} = 0.3$, and $\sin \alpha_h = 0.1, 0.05,
      0.03$. The $0.12 \, \Gamma_\mathrm{SM}
      (h^0)$ upper limit (dashed) with $ \Gamma_\mathrm{SM} (h^0) = 4$
      MeV on undetected Higgs decays~\cite{ATLAS:2024fkg}. }
\label{fig:higgsPhiPhi}
\end{figure}

The squared matrix element for the Feynman diagram shown in
Figure~\ref{fig:3:Body:Decay} reads
\begin{align}
    |\mathcal{M}(h^0 \to \varphi \, b \, \bar b \, )|^2 \approx 
        \frac{2 \mu_{\varphi \varphi h}^2 \, M_b^2 \,  M_{23}^2 \, \sin^2 \! \alpha_h}{ v_h^2 \left( M_{23}^2  - M_\varphi^2 \right)^2 }  ~~,
  \label{eq:3bodyME}
 \end{align}
where $M_b$ is the $b$ quark mass,
corrections of order $M_b/M_h$ are neglected, and
$M_{23}$ is the invariant mass of the fermion pair.
The 3-body decay width is given by
\bear
        & \Gamma (h^0 \to \varphi \, b \, \bar b \, ) \!\!\! &=  \frac{1}{2^7 \pi^3 M_h^3} 
        \int^{M_h^2}_{M_\varphi^2} \mathrm{d}M^2_{12} \int^{4 E_b E_{\bar b}}_0 \mathrm{d}M^2_{23} 
        \, |\mathcal{M}(h^0 \to \varphi \, b \, \bar b \, )|^2
  \nonumber  \\ && =
        \frac{\mu_{\varphi \varphi h}^2 \,  M_b^2 \, \sin^2 \! \alpha_h}{64 \pi^3 M_h \, v_h^2} \, R_S( M_\varphi^2/M_h^2 ) ~~,
        \label{eq:3bodyWidth}
\eear
with the function $R_S$ defined by 
\begin{align}
    R_S(x) = \frac{5 x-1}{\sqrt{1-4 x}} \, \mathrm{arcosh} \!  \left(\frac{3x -1}{2 x^{3/2} }\right) - (x-1) \left( \frac{1}{2} \ln x - 2 \right) ~~.
\end{align}
In deriving (\ref{eq:3bodyWidth}) we used a kinematic constraint on the product of $b$ quark energies: 
\begin{align}
  E_b E_{\bar b} \approx \frac{1}{4 M_{12}^2} \left( M_{12}^2 - M_\varphi^2 \right) \left( M_h^2 - M_{12}^2 \right)  ~~.
\end{align}
An equivalent expression for $\Gamma(h^0 \to \varphi \tau^+ \tau^-)$ can be obtained by substituting $M_b^2$ in (\ref{eq:3bodyME}) and (\ref{eq:3bodyWidth}) by $M_\tau^2/3$ to account for the $\tau$ mass and the sum over colors.
For $M_\varphi < M_h/2$, both scalars get on-shell and the corresponding
decay width is 
\begin{align}
    \Gamma(h^0 \rightarrow \varphi \varphi) = \frac{\mu_{\varphi \varphi h}^2}{8 \pi M_h} 
    \left(1- \frac{4 M_\varphi^2}{M_h^2} \right)^{\! 1/2}   ~~.
\end{align}
The constraint on the undetected Higgs width is 0.12 $\times$ the SM
Higgs width of $ \Gamma_\mathrm{SM} (h^0) \approx 4$
MeV~\cite{ATLAS:2024fkg}, which excludes parts of the parameter space
at small $M_\varphi $ and different choices of $M_{Z'}$ and $\sin
\alpha_h$, as shown in Figure~\ref{fig:higgsPhiPhi}.  However, for
$M_\varphi > M_h/2$, the $h^0 \rightarrow \varphi \varphi^*$ partial
width is the only new contribution to the Higgs width, and it is
negligible.

For $M_\varphi > M_h$, a decay of the SM-like Higgs into $\varphi$
opens up, which for $M_\phi < 2 M_h$ is a three body decay, and for
$M_\phi > 2 M_h$ is tree level. The decay width can be derived
analogously to the first case by changing $c_{\varphi \varphi h}
\rightarrow c_{\varphi h h }$ and $\sin \alpha_h \rightarrow \cos
\alpha_h$. However, the decay of $\varphi$ into $h^0 h^0$ is again
negligible, since for very heavy $M_\phi$ the $Z' Z'$ decay dominates.

\subsection{Mixing effects on the 125~GeV Higgs-like scalar $h^0$ and $\varphi$}
\label{subsec:mixingpheno}
If there is a sizable $\phi-h_{\text{SM}}^0$ mixing, then $\varphi$ may also be
produced by itself through gluon fusion; the cross section for this
1-loop process scales as $\sin^2\!\alpha_h$ (assuming negligible
2-loop contributions induced by the $Z'Z'\varphi$ coupling), and is
shown in Figure~\ref{fig:xsecsphi_alpha05}.  Other mixing-suppressed
production modes of $\varphi$, analogous with those of the SM Higgs,
may also be relevant. In addition, $\varphi$ may be pair produced in
SM Higgs decays: $h \to \varphi\varphi$ for $M_\varphi < M_h/2$.

The constraints for the mixing angle can be calculated from the Higgs
signal strengths $\mu_\mathrm{CMS} = 1.002 \pm 0.057$
in~\cite{CMS:2022dwd} and $\mu_\mathrm{ATLAS} = 1.05 \pm 0.06$
in~\cite{ATLAS:2022vkf}. These give a naive global average  $\mu_{\rm LHC} = 1.026 \pm 0.041$, implying
\begin{align}
    \sin \alpha_h \leq 0.24 ~~.
\label{eq:sinalphah}
\end{align}
To satisfy this bound, we will adopt the choice of $\sin \alpha_h =
0.1$, $0.05$, or $0.03$ in this work.

\subsubsection{Production modes for $\varphi$}

\begin{figure}[t!]
  \begin{center}
 \hspace*{-0.3cm} 
  \includegraphics[width=0.8\textwidth, angle=0]{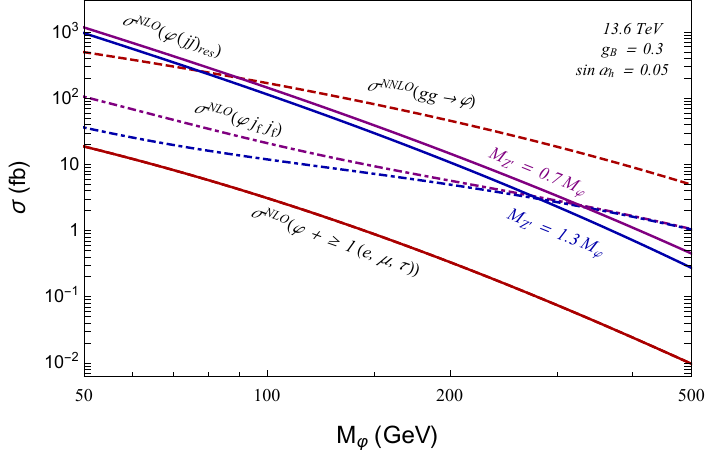}
   \caption{Cross sections at the 13.6 TeV LHC for $\varphi$ production through
     gluon fusion (dot-dashed line), in association with a $Z'$
     (solid lines), and through $Z'$ fusion (dashed lines), as a
     function of the $\varphi$ mass.  The displayed gluon fusion rate
     is calculated with MCFM~\cite{Boughezal:2016wmq,
       Campbell:2019dru} rescaled by the square of the mixing $\sin
     \alpha_h = 0.05$.  The $Z' \varphi$ and $\varphi jj$ cross
     sections are computed at NLO in $\alpha_s$ for
     $M_{Z'}/M_{\varphi}= 0.7$ (upper lines) and $1.3$ (lower lines),
     for a $Z'$ gauge coupling $g_{_B} = 0.3$; the cross section
     scales as $g_{_B}^4$ for $Z' \varphi$ production, and as
     $g_{_B}^8$ for $Z'$ fusion.
\label{fig:xsecsphi_alpha05}
 }
\end{center}
\end{figure}

In Figure~\ref{fig:xsecsphi_alpha05}, we show the cross sections for
$\phi$ production with $\sin \alpha_h = 0.05$, using
MCFM~\cite{Boughezal:2016wmq, Campbell:2019dru} for the cross sections
based on rescaling SM Higgs rates.  We can see, in comparison to the
unmixed case in Figure~\ref{fig:phixsecs_nomixing}, that the dominant
production mode becomes gluon fusion, $gg \to \varphi$, for $\varphi$
masses above 100 GeV and our specified $\sin \alpha_h = 0.05$, while
$Z'$-associated production dominates for low $\varphi$ masses.  Given
the scaling of $Z'$-associated production by $g_{_B}^4$ and the
scaling of gluon fusion by $\sin \alpha_h^2$, this crossover point
between dominant production modes is highly model-dependent.  We can
see that the next most important production mode is the sum over $Z'$-fusion and SM
vector boson fusion modes, which surpass the $Z'$-associated production modes for high
$\varphi$ masses, as already seen in
Figure~\ref{fig:phixsecs_nomixing}.  We also show the important
$\varphi \, + \geq 1 (e, \mu, \tau)$ production rates, summing over the
leptonic $W^\pm \varphi$ and $Z \varphi$ rates induced by $\sin
\alpha_h$.  Although these rates are roughly two orders of magnitude
smaller than the $Z'$-associated modes, we believe optimized analyses
for each the resonant dijet + $\varphi$ signature from $Z'$-associated
production and the lepton + $\varphi$ signature can have competitive
sensitivity, given the corresponding SM backgrounds, respectively.  An
analysis for the $Z'$-fusion mode can also reduce SM backgrounds by
focusing on the two forward jet signature.

\subsubsection{Decay modes of $\varphi$}

When $\phi-h_{\text{SM}}^0$ mixing is nonzero, the interference between
anomalon-induced processes and SM loop-induced processes is
significant, primarily reflecting the underlying chiral symmetry
respected by each sector of loop particles and fundamental spin
statistics.  We remark that even though $\phi$ is a gauge singlet, the
non-trivial interference effects evident in the $\varphi$ decay widths is
a marked departure from the simplistic mixing-angle suppressed SM
Higgs-like phenomenology that is commonly considered.

The new expression for the 1-loop decay of $\varphi$ into photons gets
contributions from the $W$ and top quark, which are suppressed by the
$\sin \alpha_h$ mixing, but also non-suppressed contributions from
anomalon loops:
\begin{flalign} 
    \Gamma (\varphi \to \gamma \gamma) &=
        \frac{\alpha^2 M_\varphi^3}{256 \pi^3} \left| \, \cos \alpha_h \, \sum_{a=1}^2 \frac{y_{aa}^{\phi (S)}}{M_a} \, F_1 \! \left( \frac{M_\varphi^2}{4 M_a^2} \right) + 
         \sin \alpha_h \sum_{a =1}^2 \frac{y_{aa}^{h_{\text{SM}} (S)}}{M_a} F_1 \!  \left( \frac{M_\varphi^2}{4 M_a^2} \right) 
             \right. \nonumber \\ & \hspace{1.8cm}
\left. +  \dfrac{\sin \alpha_h}{v_h} \left[ \, F_V \! \left( \frac{M_\varphi^2}{4 M_W^2}\right) + \dfrac{4}{3}  \, F_1\! \left( \frac{M_\varphi^2}{4 M_t^2} \right) \right] \right|^2   ~~,
\end{flalign}
where 
$y_{aa}^{\phi (S)}$ and $y_{aa}^{h_{\text{SM}} (S)}$ are given in Eqs.~(\ref{eq:ySyP}) and (\ref{eq:couplings}).  The loop functions are 
$F_1$ defined in (\ref{eq:F1}), and 
\be
F_V (x) = - 2 - \frac{3}{x} + \frac{3}{x^2} (1 - 2x) Z(x)   ~~
\ee
with $Z(x)$ given in (\ref{eq:Zx}). 
In the minimal model, the sum runs over fields $E_1$ and $E_2$. The
strength of the $\gamma \gamma$ decay depends sensitively on the
possible cancellation between the SM $W$ and $t$ loops and the charged
anomalons, mainly governed by the magnitude of $\alpha_h$ and
$v'$. The anomalons in the loop coherently add to the SM $t$, but destructively interfere with the SM $W$, resulting in a decrease of the overall strength of the diphoton
decay. Furthermore, the anomalon contribution is controlled by the
general Yukawa couplings $y^\phi$ from Eq.~(\ref{eq:couplings}), including also the fermion mixing
angles.
The new 1-loop decays into $Z' \gamma$ and $Z \gamma$, with contributions from
the anomalons which are again not suppressed by mixing, are
\begin{flalign}
\Gamma (\varphi \to Z' \gamma) &= \frac{\alpha \, M_\varphi^3 }{128 \pi^4} 
\left( 1 - \frac{M_{Z'}^2}{M_\varphi^2} \right)^{\! 3}
\left| \, \cos \alpha_h  \sum_{a,b = 1}^2  \left(  \frac{-2 c^{(Z')}_{ab}}{M_a} \right)  F^{V \gamma} \! \left(M_\varphi, M_{Z'}, M_a, M_b \right)  \right. \nonumber
 \\
& \hspace{4.2cm}  \left.  +  \sin \alpha_h  \frac{g_{_B}}{3 v_h}  \, F_2 \!
\left( \frac{4 M_t^2}{M_{\varphi}^2}, \frac{4 M_t^2}{M_{Z'}^2} \right)
\right|^2  \, ,
\end{flalign}
and
\begin{flalign}
\Gamma (\varphi \to Z \gamma) &= \frac{\alpha \, M_\varphi^3 }{128 \pi^4} 
\left( 1 - \frac{M_{Z}^2}{M_\varphi^2} \right)^{\! 3}
\left| \, \cos \alpha_h  \sum_{a,b = 1}^2  \left( \frac{-2 c^{(Z)}_{ab}}{M_a} \right) F^{V \gamma} \! \left(M_\varphi, M_{Z}, M_a, M_b \right) \right.
 \nonumber
\\
&
\hspace{0.4cm} \left. + \sin \alpha_h \frac{g }{v_h} \left[ \frac{3- 8 s_W^2 }{ 3c_W} \, F_2 \!
\left( \frac{4 M_t^2}{M_{\varphi}^2}, \frac{4 M_t^2}{M_{Z}^2} \right) \! + \frac{1}{2}  F_{V2} \! \left( \frac{4 M_W^2}{M_\varphi^2}, \frac{4 M_W^2}{M_Z^2} \right)
\right]
\right|^2  \, ,
\end{flalign}
where $c^{(Z')}_{ab} $ and $c^{(Z)}_{ab} $
are the coupling structures given by (\ref{eq:cab}), and
\begin{flalign}
F_{V2} (x, y) = c_W \left[ 4 \left( 3- \frac{s_W^2}{c_W^2} \right) I_2(x, y) + \left( \left(1 + \frac{2}{x} \right) \frac{s_W^2}{c_W^2} - \left( 5 + \frac{2}{x} \right) \right) I_1(x, y) \right] \ ,
\label{eq:FV2}
\end{flalign}
with
\begin{flalign}
I_1(x, y) = \frac{x y}{2(x - y)} + \frac{x^2 y^2}{ 2 (x-y)^2} (Z(1/x) - Z(1/y)) + \frac{x^2 y}{(x-y)^2} (g(1/x) - g(1/y))  \, ,
\label{eq:I1}
\end{flalign}
\begin{flalign}
I_2(x, y) = -\frac{x y}{2(x-y)} \left( Z(1/x) - Z(1/y) \right)  \, ,
\label{eq:I2}
\end{flalign}
where $Z(x)$ and $g(x)$ are defined in (\ref{eq:Zx}) and (\ref{eq:gx}), respectively.  We note that $F_2(x,y) = I_1(x,y) - I_2(x,y)$ from (\ref{eq:F2}), (\ref{eq:I1}), and (\ref{eq:I2}), consistent with \cite{Djouadi:2005gi}. For the $\varphi \to Z \gamma$ width, the anomalon and SM $W$ loops add together and are partially canceled by the SM $t$, in contrast to $\varphi \to \gamma \gamma$, due to the corresponding electric charges.

\begin{figure}[t!]
  \begin{center}
 \hspace*{-0.3cm} 
  \includegraphics[width=0.79\textwidth, angle=0]{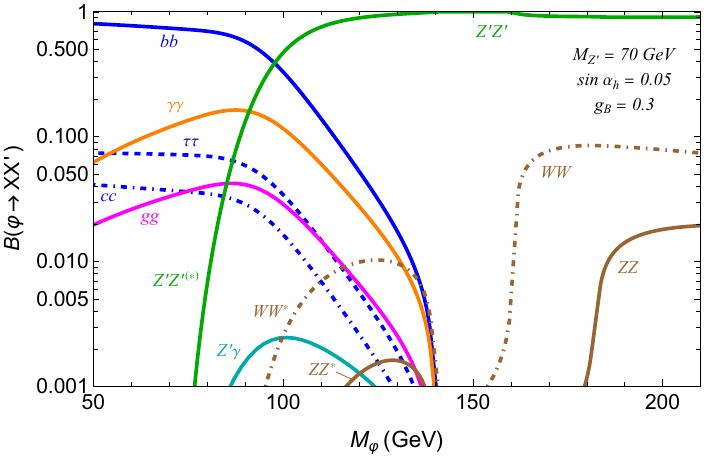}
   \vspace*{-0.2cm} 
   \caption{Branching fractions for the $U(1)_B$-breaking scalar
     $\varphi$, as a function of its mass.  The parameters fixed here
     are $\sin \alpha_h = 0.05$, $g_{_B} = 0.3$, and $M_{Z'} =
     70$~GeV.  For the loop decays $\varphi \to \gamma \gamma$, $Z'
     \gamma$, we set the anomalon masses to 200~GeV and 250~GeV, and the anomalon mixing angles are $\theta = 0.3$ and $\chi = 0.25$. \\[-0.6cm]
 \label{fig:brs_mphi_Zp70}
 }
\end{center}
\end{figure}

\begin{figure}[h!]
  \begin{center}
 \hspace*{-0.3cm} 
  \includegraphics[width=0.79\textwidth, angle=0]{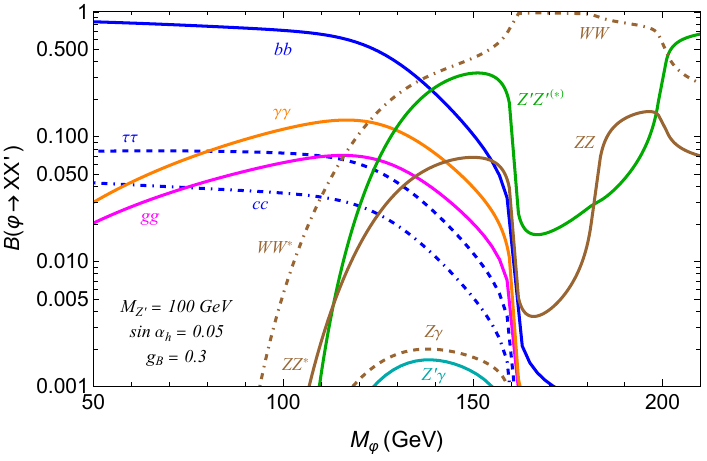}
     \vspace*{-0.2cm} 
   \caption{Branching fractions for $\varphi$ as a function of $M_\varphi$, with all the parameters fixed 
     as in Figure~\ref{fig:brs_mphi_Zp70} except for $M_{Z'} =
     100$~GeV. 
 \label{fig:brs_mphi_Zp100}
 }
\end{center}
\end{figure}

The rich variety of exotic and SM-like decay channels for $\varphi$
are shown in Figures~\ref{fig:brs_mphi_Zp70}
and~\ref{fig:brs_mphi_Zp100}.  For the mixing-induced decays to two SM
fermions, we include running of the respective Yukawa couplings using
RunDec~\cite{Chetyrkin:2000yt}.  For the $\varphi$ mass range shown,
the nominally leading decay, $\varphi \to Z' Z'^*$, is actually a
3-body process and becomes kinematically suppressed for lighter
$M_{\varphi}$.  Hence, the mixing-angle induced SM-like Higgs decay
modes compete with and are of the same order as the intrinsic
$\varphi$ decays.  Notably, the $\varphi \to \gamma \gamma$ branching
fraction is large, reflecting the anomalon-induced contribution and
the interference with the SM loops.  Similarly, the $Z' \gamma$ and $Z
\gamma$ branching fractions also exhibit different patterns compared
to the simplest models of SM gauge singlets scalars.  In general,
determining the decay patterns of $\varphi$ will give a complete
picture of the underlying $U(1)_B$ charges of the decay products.  For
$M_\varphi$ lighter than $M_{Z'}$, the two-body loop-induced $\gamma
\gamma$ decay dominates over the four-body $\varphi \to Z'{}^*Z'{}^* \to
4j$ decay.


All tree level decays into SM quarks and leptons are suppressed by
$\sin \alpha_h$, with the largest contribution coming from the decay
into two bottom quarks.  Additionally, the loop-suppressed gluon decay
is also $\sin \alpha_h$ suppressed, in contrast to
\cite{Duerr:2017whl}, since the anomalons are all colorless. For small
$\alpha_h$, the diphoton decay becomes the leading decay channel,
justifying potential searches of $\varphi$ in the diphoton spectrum.
For larger $M_\varphi$, the phase space for decays into $Z'$ and
$W$/$Z$ open up.  Here, again, the tree-level decays into SM gauge
bosons are suppressed by the Higgs mixing angle, although this can dominate over the $Z'Z'^*$ decay if the $Z'$ mass is larger than the SM electroweak gauge boson masses.  
For example, with $M_{Z'} = 100$~GeV, the $
\varphi$ dominantly decays to SM gauge bosons for the mass range $160 \lesssim M_{\varphi} \lesssim 200$~GeV.  
For $M_\varphi$ above twice the $Z'$ mass, the $Z' Z'$ decay dominates.  For scalar
masses above twice the anomalon masses, the tree-level decay of
$\varphi$ into two anomalons is allowed, which dominates similar to
the tree-level decays into vectorlike quarks in~\cite{Duerr:2017whl}.

 \begin{figure}[t]
  \begin{center}
 \hspace*{-0.3cm} 
  \includegraphics[width=0.8\textwidth, angle=0]{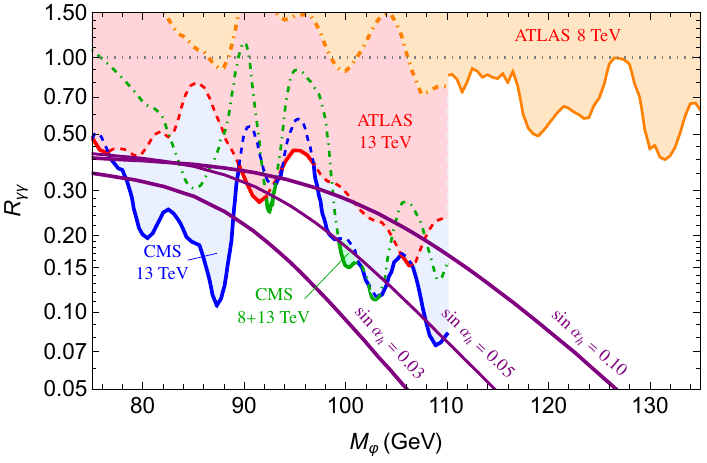}
   \caption{ Predicted $R_{\gamma \gamma}\equiv \sigma( \varphi \to
     \gamma \gamma) / \sigma( H \to \gamma \gamma)_{\text{SM}}$ as a
     function of $M_{\varphi}$ for $g_{_B} = 0.3$, $M_{Z'} = 70$~GeV,
     and $\sin\alpha_h = 0.1$, 0.05 or 0.03 (purple lines). We set the
     anomalon masses to 200~GeV and 250~GeV, and the anomalon mixing
     angles are $\theta = 0.3$ and $\chi = 0.25$.
     The shaded regions are excluded by CMS~\cite{CMS:2024yhz} and
     ATLAS~\cite{ATLAS:2024bjr, Aad:2014ioa} diphoton resonance
     searches.  }
 \label{fig:diphoton}
\end{center}
\end{figure}

The leading constraint on the $\varphi$ scalar comes from searches for
a Higgs-like scalar in the diphoton decay channel, which is
intrinsic to both unmixed scalars, $\phi$ and $h_{\text{SM}}^0$. From
Figure~\ref{fig:brs_mphi_nomixing}, for masses below $M_{Z'}$, $\phi$ dominantly
decays to two photons. Hence,
even though the introduction of a finite scalar mixing angle
$\alpha_h$ with the SM Higgs boson leads to new two-body final states
such as $b \bar{b}$, the interference of the diphoton channel for
$\varphi$ generally leads to sizeable $\varphi \to \gamma \gamma$
branching fractions, of the order of $10\%$ or larger, even until the
$2 M_{Z'}$ threshold.  As the diphoton decay enjoys a favorable signal
to background ratio for the SM Higgs discovery in the light Higgs mass
region, the considerations of an enhanced diphoton branching fraction
and good signal to background ratio support our claim that the
dominant constraint on the $\varphi$ scalar arises from reinterpreting
diphoton resonance searches.

Figure~\ref{fig:diphoton} shows the current ATLAS and CMS limits
on $R_{\gamma \gamma}$ as a function of $M_{\varphi}$, where $R_{\gamma \gamma}$ is the ratio of the signal diphoton cross section to a SM Higgs diphoton
cross section at the given $M_{\varphi}$ mass (Additional limits at
lower scalar masses are also performed in~\cite{ATLAS:2022abz}.).  We also show the model
prediction for $R_{\gamma \gamma}$ for three choices of $\sin \alpha_h
= 0.1$, $0.05$, and $0.03$.  Importantly, the diphoton signal strength
shows a nontrivial scaling dependence on $\sin \alpha_h$: in
particular, for $\sin \alpha_h = 0.1$ and $\sin \alpha_h = 0.05$, the
$R_{\gamma \gamma}$ expectation is nearly the same for $80 <
M_{\varphi} < 95$~GeV.  This is because of two competing effects.
First, as we increase $\alpha_h$, the production rate from gluon
fusion increases as $\sin^2 \alpha_h$.  On the other hand, the
interference between the SM Higgs and the $\phi$ diphoton decay
processes is destructive, because the dominant mediator for SM Higgs
to $\gamma \gamma$ is the $W$-boson, while the dominant mediators for
$\phi$ to $\gamma \gamma$ are the charged anomalons, and there is a
relative sign in their respective loop functions.  This nontrivial
scaling is a key aspect of our $Z'$ + scalar model, since the
interference of the Higgs and exotic scalar diphoton decays will be a
generic feature in chiral gauge extensions of the SM.  

We now briefly highlight the differences between the latest
experimental analyses and their corresponding sensitivities, as seen
in Figure~\ref{fig:diphoton}.  In~\cite{CMS:2024yhz} by the CMS collaboration, the kinematic
constraints on the transverse momenta are $p_T^{\gamma 1} = 30$ GeV
and $p_T^{\gamma 2} = 18$ GeV, as well as $p_T^{\gamma 1}/M_{\gamma
  \gamma} = 0.47$ and $p_T^{\gamma 2}/M_{\gamma \gamma} = 0.28$, where
$1$ and $2$ label the photon candidates with highest and
second-highest transverse momentum and $M_{\gamma \gamma}$ is the
invariant mass of the diphoton system.  A portion of the early dataset
required both photon candidates to be contained within the
electromagnetic calorimeter barrel region, $|\eta| < 1.48$, but the
remainder of the dataset includes photons covered by the endcap
regions covering $1.48 < |\eta| < 3.0$.  In~\cite{ATLAS:2024bjr} by the ATLAS collaboration, the
transverse energy requirements are $E_T > 22$ GeV for both photon
candidates and $E_T / m_{\gamma \gamma} > 22/58 \approx 0.38$, with
$|\eta| < 2.37$, excluding the transition region $1.37 < |\eta| <
1.52$.  Both analyses use boosted decision trees to further optimize
the signal selection against light jets faking photons and dielectron
background events that appear similar to photons in the calorimeter.
Further classifiers are used to characterize the production mode of
the scalar resonance according to SM Higgs production expectations,
dominantly targetting gluon fusion and vector boson fusion.  Although
the two analyses have comparable sensitivity to the $\varphi$ diphoton
signal, the improved acceptance of the diphoton system by CMS seems
most responsible for their stronger exclusion contour, which goes down
to $R_{\gamma \gamma} \simeq 0.07$ around $M_{\varphi} = 109$~GeV,
while the best constraint from ATLAS is down to $R_{\gamma \gamma}
\simeq 0.15$ for $M_{\varphi} = 106$~GeV.

We also show the earlier $\sqrt{s} = 8$~TeV analysis from
ATLAS~\cite{Aad:2014ioa}, which uses similar cuts as the newer
exclusion papers as well as the Higgs discovery papers.  We advocate
new analyses that cover the diphoton invariant mass region above
110~GeV using current and future datasets.  The possible overlap with
the 125~GeV Higgs diphoton signal can be either masked with a narrow
window, or more robustly, the mixing between the $\varphi$ and the
125~GeV Higgs boson can be treated phenomenologically following the
formalism in~\cite{LoChiatto:2024guj, Frank:2006yh, Fuchs:2016swt}.

 \begin{figure}[tbh!]
  \begin{center}
 \hspace*{-0.3cm} 
  \includegraphics[width=0.8\textwidth, angle=0]{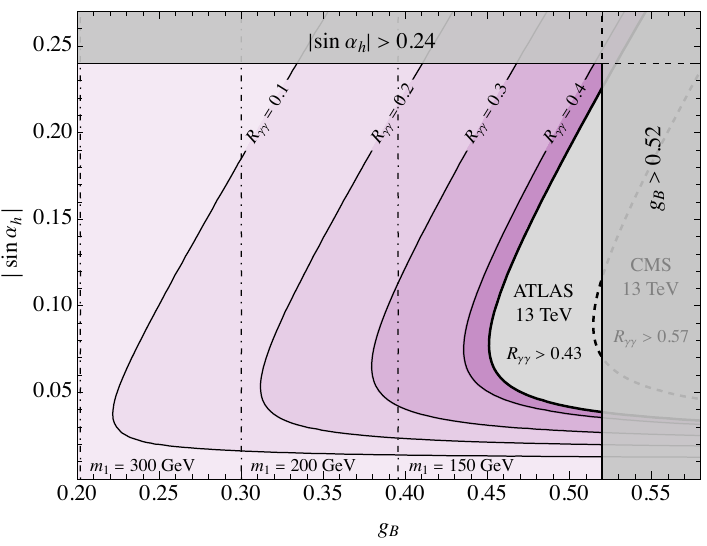}
   \caption{Contours of $R_{\gamma \gamma}$ for the $\varphi$ diphoton
     signal as a function of $g_{_B}$ and $|\sin \alpha_h|$, with
     $M_\varphi = 95.4$~GeV and $M_{Z'} = 70$~GeV.  The ATLAS
     constraint (solid black) of $R_{\gamma \gamma} <
     0.43$~\cite{ATLAS:2024bjr} and CMS constraint (dashed black) of
     $R_{\gamma \gamma} < 0.57$~\cite{CMS:2024yhz} are also shown as
     gray regions.  The vertical dot-dashed lines indicate the lighter
     $m_1$ charged anomalon mass.  We also show the excluded region
     with $|\sin \alpha_h| > 0.24$ from the 125~GeV Higgs signal
     strength combination as well as $g_B >
     0.52$~\cite{Dobrescu:2021vak} from dijet resonance searches.  }
\label{fig:excessFixedMZp}
\end{center}
\end{figure}

We remark that the existing constraints from diphoton searches can
also be improved if a dedicated dijet + diphoton search is performed,
given the expectation for $Z'$-associated production rates leading to
an associated dijet resonance.

Interestingly, both experiments show an upward fluctuation in their
$R_{\gamma \gamma}$ constraints for a diphoton resonance around
95~GeV, which could be explained by $\varphi$\footnote{Related
  explanations can be found in~\cite{Fox:2017uwr}.}.  Namely, at
95.4~GeV, CMS observes a $2.9\sigma$ local ($1.3\sigma$ global)
excess~\cite{CMS:2024yhz} while ATLAS only observes a $1.7\sigma$
local significance~\cite{ATLAS:2024bjr}.  In
Figure~\ref{fig:excessFixedMZp}, we show the contours of $R_{\gamma
  \gamma}$ as a function of $g_B$ and $|\sin \alpha_h|$ for $M_\varphi
= 95.4$~GeV and $M_{Z'} = 70$~GeV.  The ATLAS and CMS measurements
exclude $R_{\gamma \gamma} > 0.43$ and $R_{\gamma \gamma} > 0.57$,
respectively.  We also show the corresponding $M_1$ charged anomalon
mass as vertical dot-dashed lines, and additional exclusions on $|\sin
\alpha_h| > 0.24$ from (\ref{eq:sinalphah}) and $g_B > 0.52$ from
Ref.~\cite{Dobrescu:2021vak}.  We note that the $\tau \tau$ excess from CMS~\cite{CMS:2022goy} 
requires an order one ratio between $\tau \tau$ and $\gamma \gamma$ widths, which
cannot be accommodated in our model.

\section{Conclusions}
\setcounter{equation}{0}
\label{sec:conclusions}

In this work, we have studied the phenomenology of the scalar $\phi$ associated with the spontaneous breaking of a $U(1)$ gauge symmetry beyond the SM. The generic mechanisms for $\phi$ production at hadron colliders, which only require $U(1)$ charges for light SM quarks,
 are in association with a $Z'$ boson, or through $Z'$ boson fusion (see diagrams in Figure~\ref{fig:phidiags}). Our results for the cross sections of these processes at the LHC, computed at NLO,  are shown in Figure~\ref{fig:phixsecs_nomixing}.
 
If the SM quarks carry charges under the new $U(1)$, then the cancellation of the gauge anomalies typically requires the existence of new fermions (anomalons) with chiral charges under both the SM gauge groups and $U(1)$ symmetry.
We emphasize that, like in the SM, the underlying chiral
structure of the $U(1)$ symmetry is characterized by a single vev,
and hence the $Z'$, $\phi$, and anomalon masses are unlikely to 
arbitrarily decouple from each other without violating perturbative unitarity.

The anomalons induce $\phi$ decays, at one loop, 
into $\gamma \gamma$, $Z' \gamma$, $Z \gamma$, $ZZ$, and $WW$, with the $Z$ and $W$ bosons being on- or off-shell depending on the $\phi$ mass.
Another generic decay mode of $\phi$, which occurs at tree level but may be highly phase-space suppressed, is into a pair of $Z'$ bosons with one or both of them being off-shell depending on the mass ratio $M_\phi /M_{Z'}$. 

In the presence of the portal coupling of $\phi$ to the SM Higgs doublet, the mass mixing between $\phi$ and $h^0_{\rm SM}$ gives rise to a physical scalar, $\varphi$, in addition to the already discovered Higgs boson $h^0$. This mixing induces 
additional $\varphi$ decays (for example, $\varphi \to b\bar b, \tau^+\tau^-$) and 
production mechanisms (gluon fusion, associated $\varphi W$, etc.). In addition, trilinear scalar interactions lead to $\varphi \to h^0 h^{0 (*)}$ or
$h^0 \to \varphi \varphi^{(*)}$, depending on the $\varphi$ mass.

A particular challenging scenario to disentangle at colliders is that where the $Z'$ is leptophobic, and the anomalons are heavier than $M_{Z'}/2$.
To obtain a quantitative understanding of the various LHC signals in that scenario, we focused on the gauged quark-universal $U(1)_B$ symmetry, in which all SM quarks have the same coupling to the $Z'$. Furthermore, to compare the various scalar decay modes, we considered the minimal set of anomalons in Section~\ref{sec:theory}, and showed that there is an interesting interplay between the $\phi \to Z'Z^{\prime (*)}$,  $\gamma \gamma$, $Z' \gamma $, $Z \gamma$, $ZZ$ and $WW$ branching fractions (see Figure~\ref{fig:brs_mphi_nomixing}).
Various aspects of the richer phenomenology of $\varphi$, in the presence of a sizable $\phi-h^0_{\rm SM}$ mixing, are analyzed in Section~\ref{sec:phiHiggs}.

Although the anomalons could be difficult to probe directly because of
the electroweak production cross sections and model-dependent decays,
the gauge singlet $\phi$ is promising since the
anomalons mediate a nondecoupling diphoton decay.  Hence, if the
direct $\phi \to Z' Z'$ decay is kinematically forbidden and $\phi \to
Z' jj$ is then suppressed by three-body phase space, the $\gamma
\gamma$ and $Z' \gamma$ decays of $\phi$ are attractive discovery
channels.  When we include possible mixing with the SM Higgs field, the loop-induced decays gain additional impetus and can
reveal new interference patterns of the underlying Yukawa couplings.

If both a $Z'$ boson and a $\varphi$ scalar will be discovered, there are many possible measurements to be made, including associated $Z' \varphi$ production and  $Z'$ fusion followed by several $\varphi$ decay modes shown in Figures~\ref{fig:brs_mphi_Zp70} and~\ref{fig:brs_mphi_Zp100} (or in Figure~\ref{fig:brs_mphi_nomixing} if the mixing with the SM Higgs is negligible).  That set of measurements would elucidate the properties of the bosons and of the anomalons, as well as the new gauge structure.

\bigskip\bigskip

\noindent{\it Acknowledgments: } 
This work is supported by the Cluster of Excellence {\em Precision
  Physics, Fundamental Interactions and Structure of Matter\/}
(PRISMA${}^+$ -- EXC~2118/1) within the German Excellence Strategy
(project ID 390831469).  LA and FY would like to thank the Theoretical
Physics Division at Fermilab for their hospitality and support while
the work was completed.  
Fermilab is administered by Fermi Forward Discovery Group, LLC under Contract No. 89243024CSC000002 with the U.S. Department of Energy, Office of Science, Office of High Energy Physics.

\bigskip   
   
\providecommand{\href}[2]{#2}\begingroup\raggedright

\vfil
\end{document}